\documentclass[journal,comsoc]{IEEEtran}

\usepackage[T1]{fontenc}
\usepackage{amsmath}
\usepackage[cmintegrals]{newtxmath}
\usepackage{multirow}
\usepackage{array}
\usepackage{color}
\usepackage{listings}
\usepackage{fancyvrb}
\usepackage{booktabs}
\usepackage{cuted}
\usepackage{inputenc}
\usepackage[pdftex]{graphicx}
\usepackage{placeins}
\usepackage{amsmath, amssymb, tabularx}
\usepackage{lipsum}
\usepackage{hyperref}
\usepackage{tabto}
\usepackage{setspace}
\usepackage[lined,algonl,boxed]{algorithm2e}
\usepackage{balance}
\usepackage{enumitem}
\usepackage{rotating}
\usepackage{pdflscape}
\usepackage{pdfcomment}

\begin{document}
\title{Resource Management in Fog/Edge Computing: A Survey}


\author{Cheol-Ho~Hong~and~Blesson~Varghese~
\thanks{Cheol-Ho Hong is with the School of Electrical and Electronics Engineering, Chung-Ang University, South Korea. E-mail: cheolhohong@cau.ac.kr}
\thanks{Blesson Varghese is with the School of Electronics, Electrical Engineering and Computer Science, Queen's University Belfast, United Kingdom. E-mail: b.varghese@qub.ac.uk}
}

\maketitle

\thispagestyle{plain}
\pagestyle{plain}

\begin{abstract}
Contrary to using distant and centralized cloud data center resources, employing decentralized resources at the edge of a network for processing data closer to user devices, such as smartphones and tablets, is an upcoming computing paradigm, referred to as fog/edge computing. Fog/edge resources are typically resource-constrained, heterogeneous, and dynamic compared to the cloud, thereby making resource management an important challenge that needs to be addressed. This article reviews publications as early as 1991, with 85\% of the publications between 2013--2018, to identify and classify the architectures, infrastructure, and underlying algorithms for managing resources in fog/edge computing.
\end{abstract}


\IEEEpeerreviewmaketitle

\section{Introduction}
\label{sec:introduction}
Accessing remote computing resources offered by cloud data centers has become the de facto model for most Internet-based applications.
Typically, data generated by user devices such as smartphones and wearables, or sensors in a smart city or factory are all transferred to geographically distant clouds to be processed and stored.
This computing model is not practical for the future because it is likely to increase communication latencies when billions of devices are connected to the Internet~\cite{nextgen-1}.
Applications will be adversely impacted because of the increase in communication latencies, thereby degrading the overall Quality-of-Service (QoS) and Quality-of-Experience (QoE).

\begin{figure}
	\centering
	\includegraphics[width=0.5\textwidth]{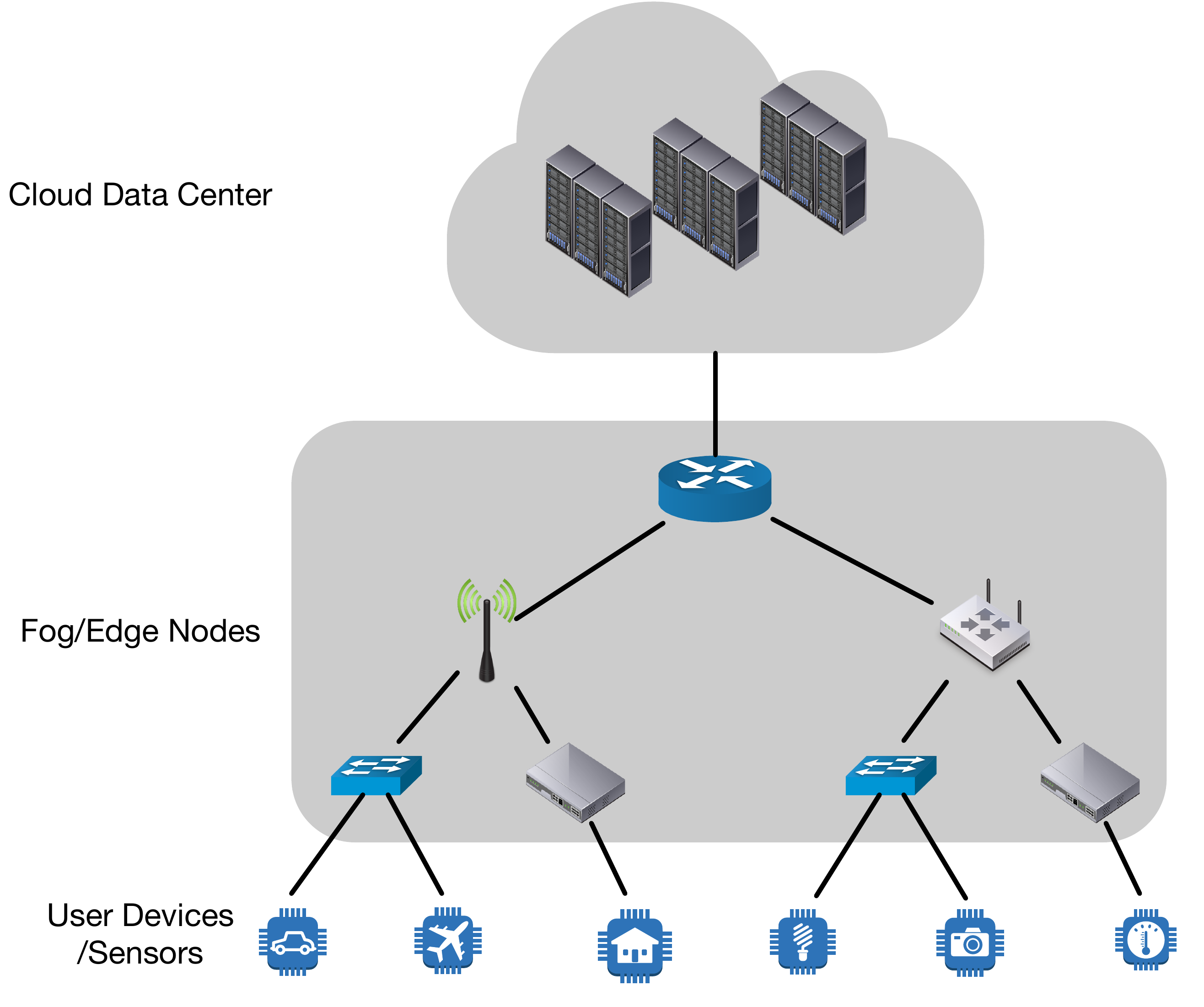}
	\caption{A fog/edge computing model comprising the cloud, resources at the edge of the network, and end-user devices or sensors}
	\label{fig:fog_computing}
\end{figure}

\begin{sidewaysfigure*}
\includegraphics[width=\textwidth,height=\textheight,keepaspectratio]{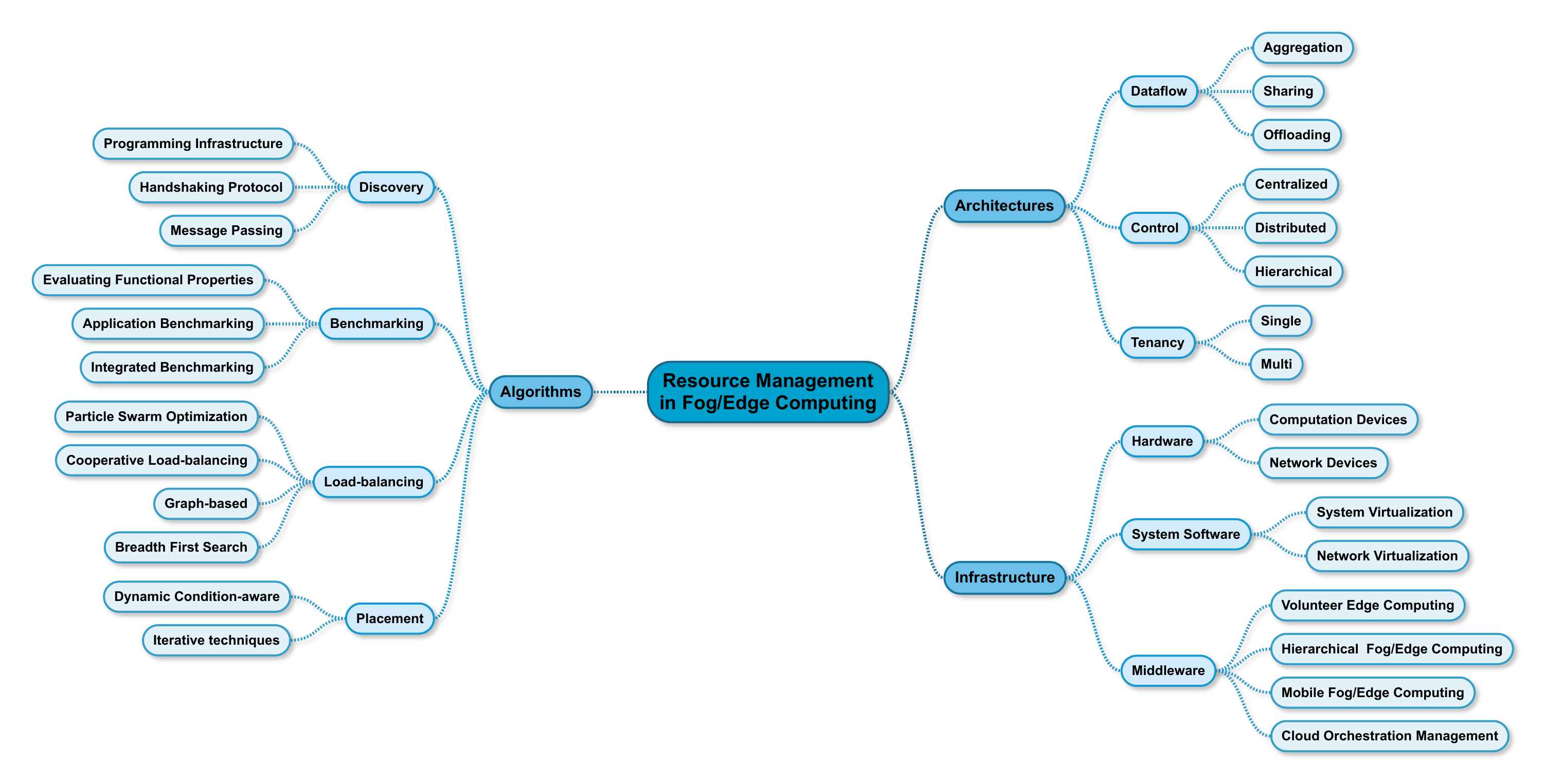}
    \caption{A classification of the architectures, infrastructure, and algorithms for resource management in fog/edge computing}
    \label{fig:overall}
\end{sidewaysfigure*}

An alternative computing model that can alleviate the above problem is bringing computing resources closer to user devices and sensors, and using them for data processing (even if only partial)~\cite{disc-1, intro-2}.
This would reduce the amount of data sent to the cloud, consequently reducing communication latencies.
To realize this computing model, the current research trend is to decentralize some of the computing resources available in large data centers by distributing them towards the edge of the network closer to the end-users and sensors, as depicted in Figure \ref{fig:fog_computing}.
These resources may take the form of either (i) dedicated `micro' data centers that are conveniently and safely located within public/private infrastructure or (i) Internet nodes, such as routers, gateways, and switches that are augmented with computing capabilities. A computing model that makes use of resources located at the edge of the network is referred to as `edge computing'~\cite{satyanarayanan2009case, intro-3}. A model that makes use of both edge resources and the cloud is referred to as `fog computing'~\cite{intro-4, intro-5}.

Contrary to cloud resources, the resources at the edge are: (i) resource constrained - limited computational resources because edge devices have smaller processors and a limited power budget, (ii) heterogeneous - processors with different architectures, and (iii) dynamic – their workloads change, and applications compete for the limited resources. Therefore, managing resources is one of the key challenges in fog and edge computing. The focus of this article is to review the architectures, infrastructure, and algorithms that underpin resource management in fog/edge computing. Figure~\ref{fig:overall} presents the areas covered by this article.



\begin{figure}
	\centering
	\includegraphics[width=0.48\textwidth]{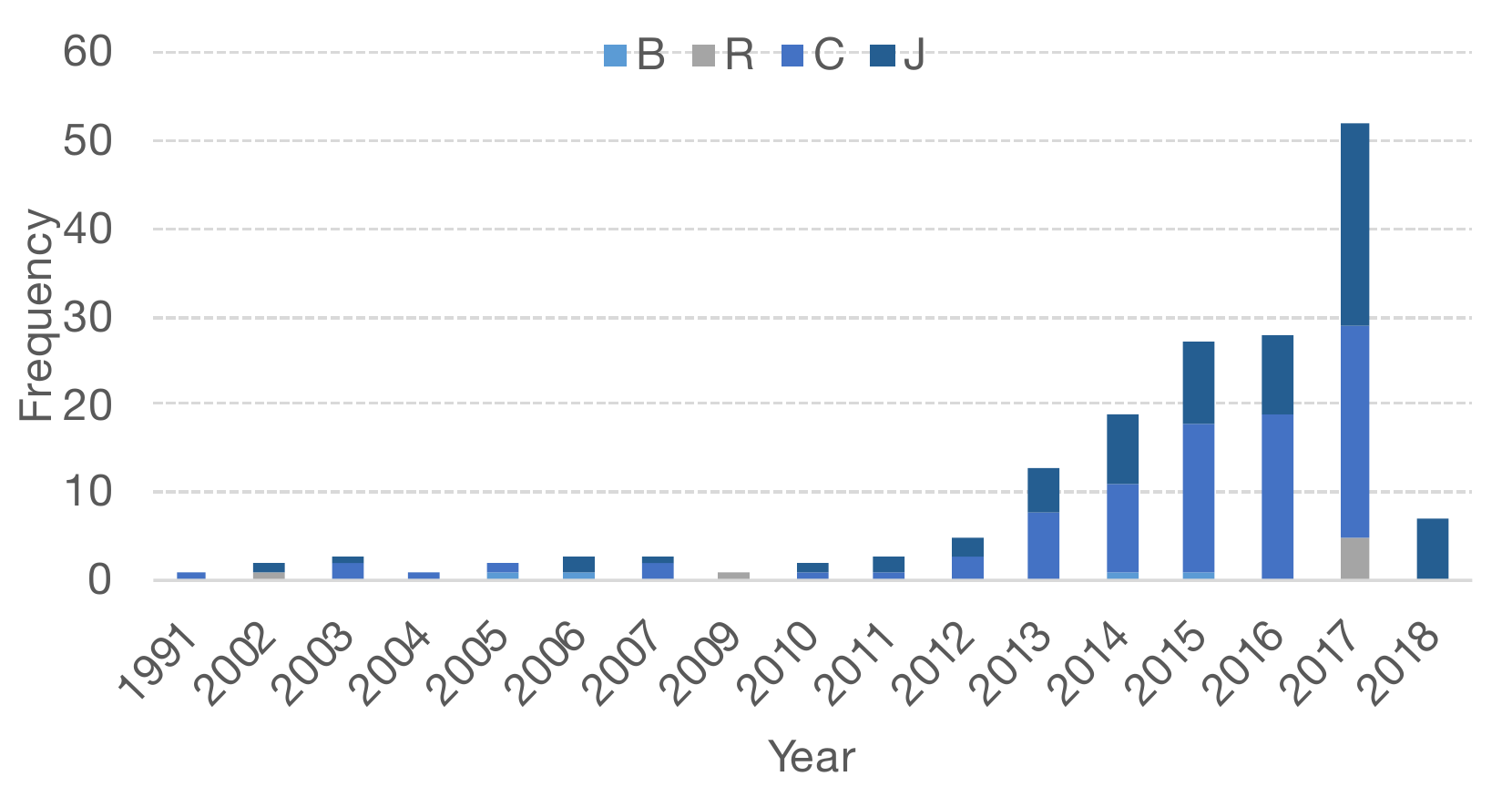}
	\caption{A histogram of the total number of research publications on resource management in fog/edge computing reviewed by this article. Legend: B - books or book chapters; R - reports, including articles available on pre-print servers or white papers; C - conference or workshop papers; J - journal or magazine articles.}
	\label{fig:histogram-overall}
\end{figure}

Figure~\ref{fig:histogram-overall} shows a histogram of the total number of research publications reviewed by this article between 1991 and 2018 under the categories: (i) books and book chapters, (ii) reports, including articles available on pre-print servers or white papers, (iii) conference or workshop papers, and (iv) journal or magazine articles. Similar histograms are provided for each section. More than 85\% of the articles reviewed were published from 2013.

The remainder of this article is structured as follows. Section~\ref{sec:architecture} discusses resource management architectures, namely the dataflow, control, and tenancy architectures. Section~\ref{sec:infrastructure} presents the infrastructure used for managing resources, such as the hardware, system software, and middleware  employed. Section~\ref{sec:algorithms} highlights the underlying algorithms, such as discovery, benchmarking, load balancing, and placement. Section~\ref{sec:conclusions} suggests future directions and concludes the paper.

\section{Architectures}
\label{sec:architecture}
In this survey, the architectures used for resource management in fog/edge computing are classified on the basis of data flow, control, and tenancy.

\begin{itemize}[leftmargin=0.3cm]
\item \textit{Data flow architectures}: These architectures are based on the direction of movement of workloads and data in the computing ecosystem. For example, workloads could be transferred from the user devices to the edge nodes or alternatively from cloud servers to the edge nodes.

\item \textit{Control architectures}: These architectures are based on how the resources are controlled in the computing ecosystem. For example, a single controller or central algorithm may be used for managing a number of edge nodes. Alternatively, a distributed approach may be employed.

\item \textit{Tenancy architecture}: These architectures are based on the support provided for hosting multiple entities in the ecosystem. For example, either a single application or multiple applications could be hosted on an edge node.
\end{itemize}

\begin{figure}
	\centering
	\includegraphics[width=0.48\textwidth]{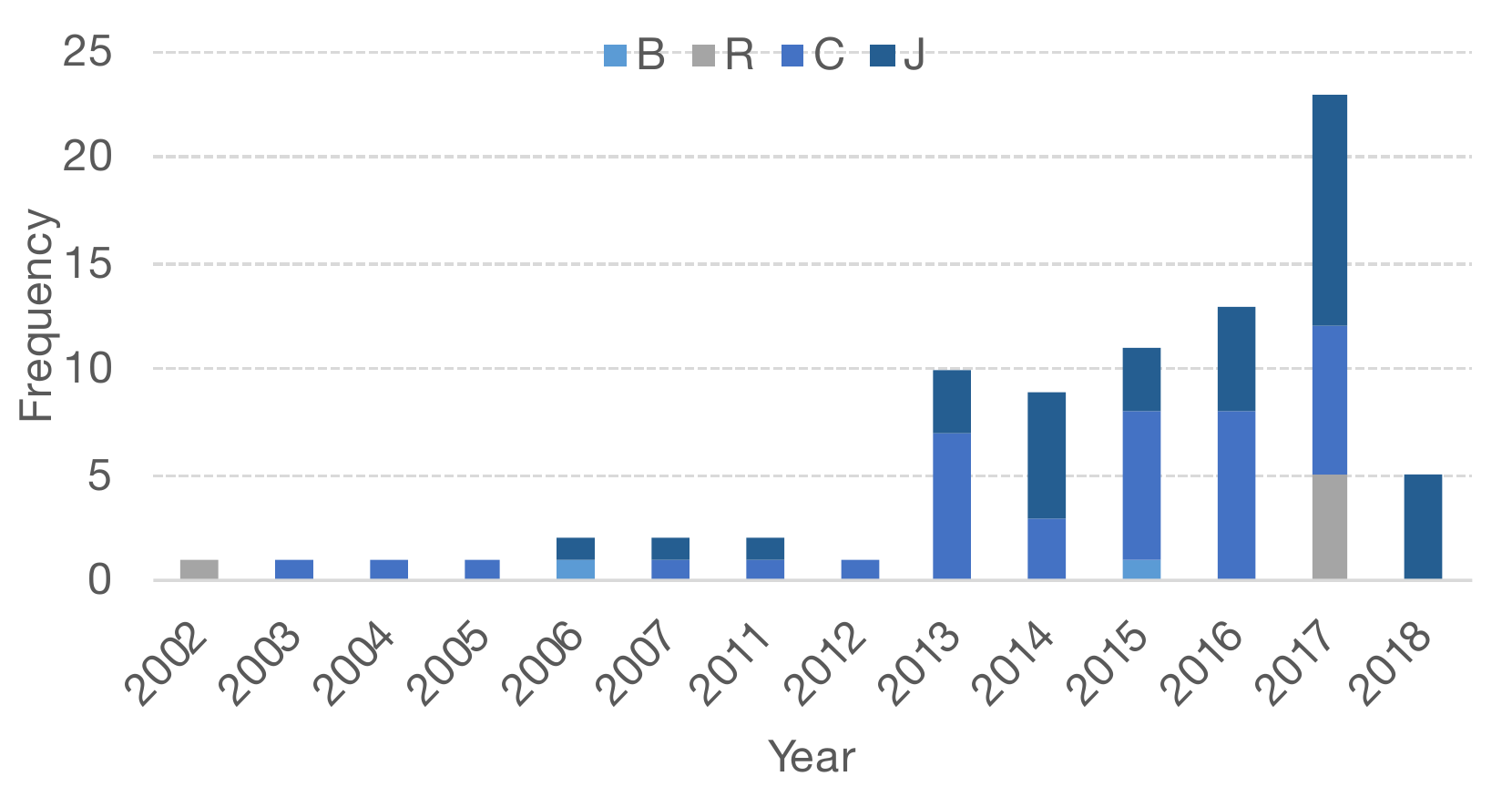}
	\caption{A histogram of publications reviewed for the classification of architectures for resource management in fog/edge computing. Legend: B - books or book chapters; R - reports, including articles available on pre-print servers or white papers; C - conference or workshop papers; J - journal or magazine articles.}
	\label{fig:histogram-architecture}
\end{figure}

The survey used 82 research publications to obtain the classification of the architectures shown in the histogram in Figure~\ref{fig:histogram-architecture}. 86\% of publications have been published since 2013.

\subsection{Data Flow}
This survey identifies key data flow architectures based on how data or workloads are transferred within a fog/edge computing environment. This section considers three data flow architectures, namely aggregation, sharing, and offloading.

\subsubsection{\textbf{Aggregation}}
\label{sec:aggregation}
\begin{figure}
	\centering
	\includegraphics[width=0.5\textwidth]{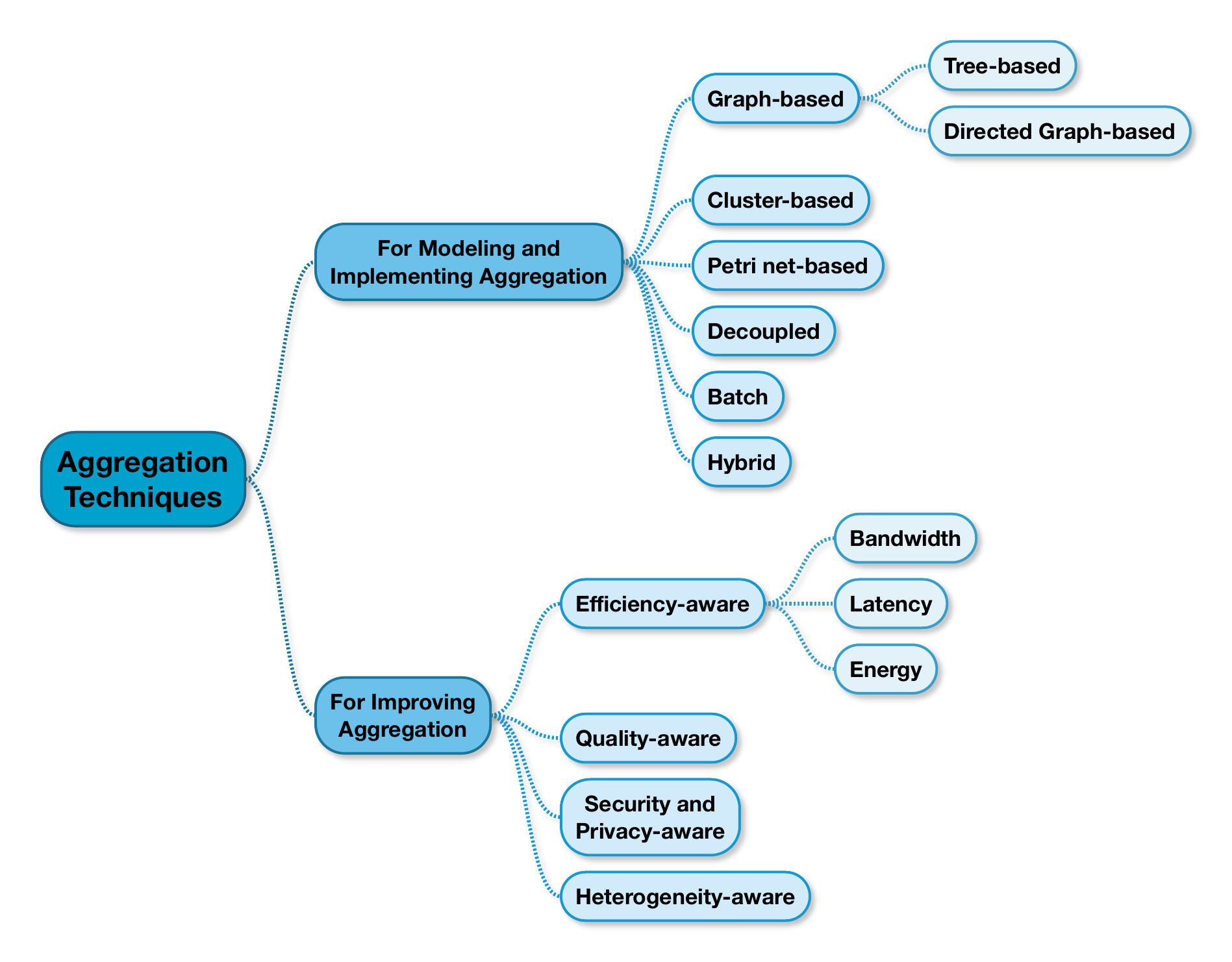}
	\caption{A classification of aggregation techniques}
	\label{fig:aggregation}
\end{figure}

In the aggregation model, an edge node obtains data generated from multiple end devices that is then partially computed for pruning or filtering. The aim in the aggregation model is to reduce communication overheads, including preventing unnecessary traffic from being transmitted beyond the edge of the network. Research on aggregation can broadly be classified on the basis of (i) techniques for modeling and implementing aggregation, and (ii) techniques for improving aggregation, as shown in Figure~\ref{fig:aggregation}.

\textit{i. Techniques for Modeling and Implementing Aggregation}: The underlying techniques implemented for supporting aggregation have formed an important part of Wireless Sensor Networks (WSNs)~\cite{agg-1} and distributed data stream processing~\cite{agg-2}. Dense and large-scale sensor networks cannot route all data generated from sensors to a centralized server, but instead need to make use of intermediate nodes along the data path that aggregate data. This is referred to as in-network data aggregation~\cite{agg-3}. We consider WSNs to be predecessors of modern edge computing systems. Existing research in the area of in-network data aggregation can be classified into the following six ways on the basis of the underlying techniques used for modeling and implementing aggregation:

\textit{a. Graph-based Techniques}: In this survey, we report two graph-based techniques that are used for data aggregation, namely tree-based and directed graph-based techniques.

\textit{Tree-based Techniques}: Two examples of tree-based techniques are Data Aggregation Trees (DATs) and spatial index trees. DATs are commonly used for aggregation in WSNs using Deterministic Network Models (DNMs) or Probabilistic Network Models (PNMs). Recent research highlights the use of PNMs over DNMs for making realistic assumptions of lossy links in the network by using tree-based techniques for achieving load balancing~\cite{agg-4}. Spatial index trees are employed for querying within networks, but have recently been reported for aggregation. EGF is an energy efficient index tree used for both data collection and aggregation~\cite{agg-5}. This technique is demonstrated to work well when the sensors are unevenly distributed. The sensors are divided into grids, and an index tree is first constructed. Based on the hierarchy, an EGF tree is constructed by merging neighboring grids. Multi-region queries are aggregated in-network and then executed.

\textit{Directed Graph-based Techniques}: The Dataflow programming model uses a directed graph and is used for WSN applications. Recently, a Distributed Dataflow (DDF) programming model has been proposed in the context of fog computing~\cite{agg-6}. The model is based on the MQTT protocol, supports the deployment of flow on multiple nodes, and assumes the heterogeneity of devices~\cite{agg-12}.

\textit{b. Cluster-based Techniques}: These techniques rely on clustering the nodes in the network. For example, energy efficiency could be a key criterion for clustering the nodes. One node from each cluster is then chosen to be a cluster head. The cluster head is responsible for local aggregation in each cluster and for transmitting the aggregated data to another node. Clustering techniques for energy efficient data aggregation have been reported~\cite{agg-7}. It has been highlighted that the spatial correlation models of sensor nodes cannot be used accurately in complex networks. Therefore, Data Density Correlation Degree (DDCD) clustering has been proposed~\cite{agg-8}.

\textit{c. Petri Net-based Techniques}: In contrast to tree-based techniques, recent research highlights the use of High Level Petri Net (HLPN) referred to as RedEdge for modeling aggregation in edge-based systems~\cite{agg-9}. Given that fog/edge computing accounts for three layers, namely the cloud, the user device, and the edge layers, techniques that support heterogeneity are required. HLPN facilitates heterogeneity, and the model is validated by verifying satisfiability using an automated solver. The data aggregation strategy was explored for a smart city application and tested for a variety of efficiency metrics, such as latency, power, and memory consumption.

\textit{d. Decoupled Techniques}: The classic aggregation techniques described above usually exhibit high inaccuracies when data is lost in the network. The path for routing data is determined on the basis of the aggregation technique. However, Synopsis Diffusion (SD) is a technique proposed for decoupling routing from aggregation so that they can be individually optimized to improve accuracy~\cite{agg-10}. The challenge in SD is that if one of the aggregating nodes is compromised, false aggregations will occur. More recently, there has been research to filter outputs from compromised nodes~\cite{agg-11}. In more recent edge-based systems, Software-Defined Networking (SDN) is employed to decouple computing from routing~\cite{agg-12, agg-13}. SDN will be considered in Section \ref{Network Virtualization}.

\textit{e. Batch Techniques}: This model of aggregation is employed in data stream processing. The data generated from a variety of sources is transmitted to a node where the data is grouped at time intervals to a batch job. Each batch job then gets executed on the node. For example, the underlying techniques of Apache Flink rely on batch processing of incoming data\footnote{https://flink.apache.org/}. Similarly, Apache Spark\footnote{https://spark.apache.org/} employs the concept of Discretized Streams (or D-Streams)~\cite{agg-14}, a micro-batch processing technique that periodically performs batch computations over short time intervals.

\textit{f. Hybrid Techniques}: These techniques combine one or more of the techniques considered above. For example, the Tributary-Delta approach combines tree-based and Synopsis Diffusion (SD) techniques in different regions of the network~\cite{agg-15}. The aim is to provide low loss rate and present few communication errors while maintaining or improving the overall efficiency of the network.

\textit{ii. Techniques for Improving Aggregation}: Aggregation can be implemented, such that it optimizes different objectives in the computing environment. These objectives range from communication efficiency in terms of bandwidth, latency, and energy constraints (that are popularly used) to the actual quality of aggregation (or analytics) that is performed on the edge node. The following is a classification obtained after surveying existing research on techniques for improving aggregation:

\textit{a. Efficiency-aware Techniques}: We present three categories of efficiency-aware techniques: the first for optimizing bandwidth, the second for minimizing latency, and the third for reducing energy consumption.

\textit{Bandwidth-aware}: The Bandwidth Efficient Cluster-based Data Aggregation (BECDA) algorithm has three phases~\cite{agg-16}. First, distributed nodes are organized into a number of clusters. Then, each cluster elects a cluster head that aggregates data from within the cluster. Thereafter, each cluster head contributes to intra-cluster aggregation. This approach utilizes bandwidth efficiently for data aggregation in a network and is more efficient than predecessor methods.

\textit{Latency-aware}: Another important metric that is often considered in edge-based systems for aggregation includes latency~\cite{agg-17, agg-18}. A mediation architecture has been proposed in the context of data services for reducing latency~\cite{agg-19}. In this architecture, policies for filtering data produced by the source based on concepts of complex event processing are proposed. In the experimental model, requests are serviced in near real-time with minimum latency. There is a trade-off against energy efficiency when attempting to minimize latency~\cite{agg-20}. Therefore, techniques to keep latency to a minimum while maintaining constant energy consumption were employed.

\textit{Energy-aware}: Research in energy efficiency of data aggregation focuses on reducing the power consumption of the network by making individual nodes efficient via hardware and software techniques. For example, in a multi-hop WSN, the energy consumption trade-off with aggregation latency has been explored under the physical interference model~\cite{agg-20}. A successive interference cancellation technique was used, and an energy efficient minimum latency data aggregation algorithm proposed. The algorithm achieves lower bounds of latency while maintaining constant energy. In a mobile device-based edge computing framework, RedEdge, it was observed that the energy consumption for data transfer was minimized~\cite{agg-9}. However, there is a data processing overhead on the edge node. Energy awareness techniques for edge nodes are an open research area\footnote{http://www.uniserver2020.eu/}.

\textit{b. Quality-aware Techniques}: Selective forwarding is a technique in which data from end devices are conditionally transmitted to a node for reducing overheads. `Quality-aware' in this context refers to making dynamic decisions for improving the quality of predictive analytics in selective forwarding~\cite{agg-21}. In a recent study, the optimal stopping theory was used for maximizing the quality of aggregation without compromising the efficiency of communication~\cite{agg-22}. It was noted that instantaneous decision-making that is typically employed in selective forwarding does not account for the historical accuracy of prediction. Quality awareness is brought into this method by proposing optimal vector forwarding models that account for historical quality of prediction.

\textit{c. Security-aware Techniques}: Aggregation occurring at an edge node between user devices and a public cloud needs to be secure and ensure identity privacy. An Anonymous and Secure Aggregation (ASAS) scheme~\cite{agg-23} in a fog environment using elliptic curve public-key cryptography, bilinear pairings, and a linearly holomorphic cryptosystem, namely the Castagnos-Laguillaumie cryptosystem~\cite{agg-24}, has been developed. Another recently proposed technique includes the Lightweight Privacy-preserving Data Aggregation (LPDA) for fog computing~\cite{agg-25}. LPDA, contrary to ASAS, is underpinned by the homomorphic Paillier encryption, the Chinese Remainder Theorem, and one-way hash chain techniques. Other examples of privacy-aware techniques include those employed in fog computing-based vehicle-to-infrastructure data aggregation~\cite{agg-25a}.

\textit{d. Heterogeneity-aware Techniques}: Edge-based environments are inherently heterogeneous~\cite{agg-26}. Traditional cloud techniques for data aggregation have assumed homogeneous hardware, but there is a need to account for heterogeneity. Some research takes heterogeneous nodes into account for data aggregation in WSNs~\cite{agg-27, agg-28}. Heterogeneous edge computing is still in infancy~\cite{agg-29}.

\subsubsection{\textbf{Sharing}}
\label{sec:sharing}
Contrary to the aggregation model, the sharing model is usually employed when the workload is shared among peers. This model aims at satisfying computing requirements of a workload on a mobile device without offloading it into the cloud, but onto peer devices that are likely to be battery-powered. This results in a more dynamic network given that devices may join and leave the network without notice. Practically feasible techniques proposed for cooperative task execution will need to be inherently energy aware. Research in this area is generally pursued under the umbrella of Mobile Cloud Computing (MCC)~\cite{sharing-1} and Mobile Edge Computing (MEC)~\cite{sharing-2} and is a successor to peer-to-peer computing~\cite{sharing-3}.

\begin{figure}
	\centering
	\includegraphics[width=0.5\textwidth]{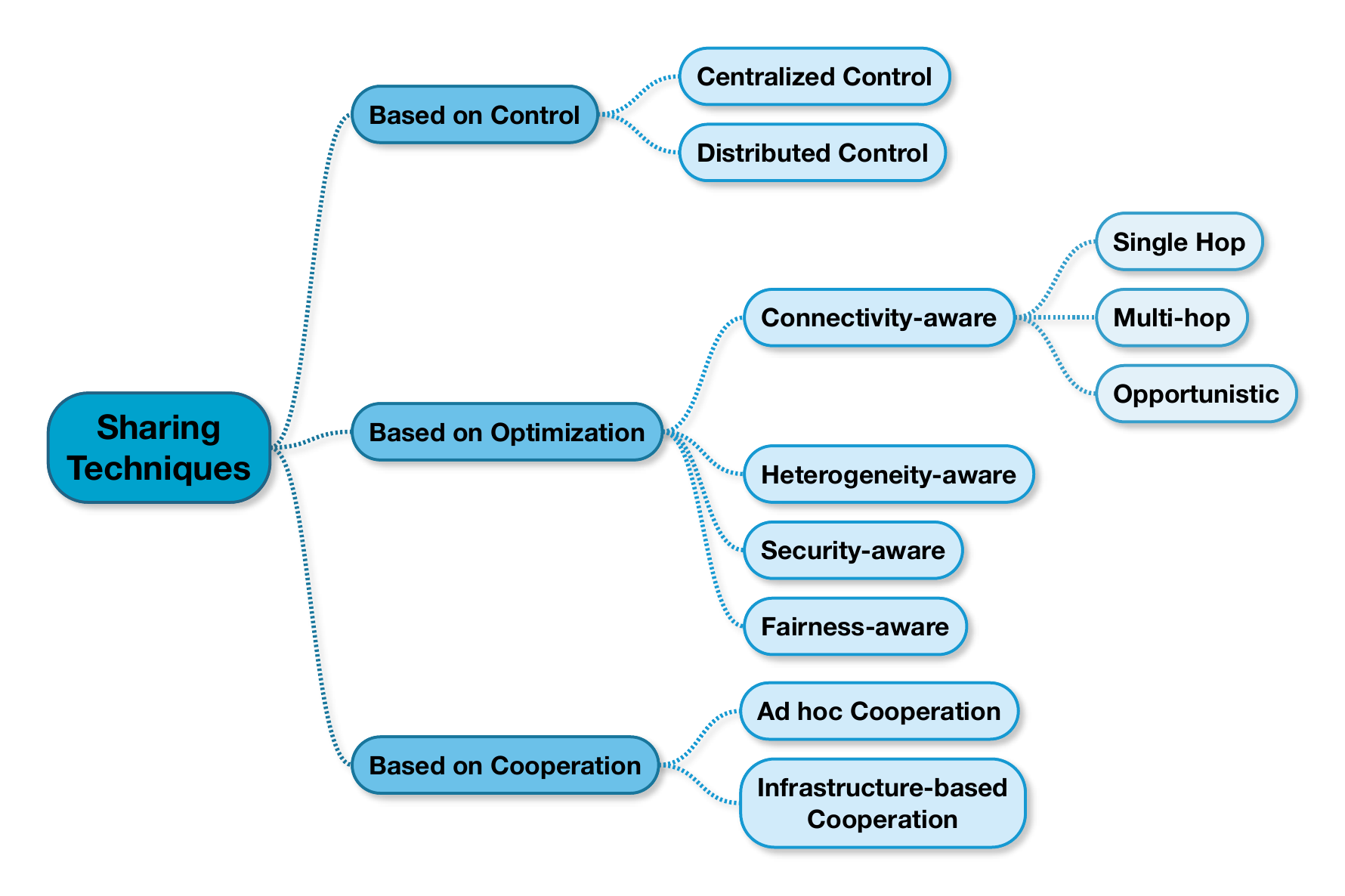}
	\caption{A classification of sharing techniques}
	\label{fig:sharing}
\end{figure}

Research on techniques for sharing can be classified into the following three ways, as shown in Figure~\ref{fig:sharing}:

\textit{i. Based on Control}: Research on control of the sharing model employed in mobile edge devices can be distinguished on the basis of (a) centralized control and (b) distributed control.

\textit{a. Centralized Control}: In this technique, a centralized controller is used to manage the workload on each edge device in a network. For example, a collection of devices at the edge is modeled as a Directed Acyclic Graph (DAG)-based workflow. The coordination of executing tasks resides with a controller in the cloud~\cite{sharing-4, sharing-5}. A Software Defined Cooperative Offloading Model (SDCOM) was implemented based on Software Defined Networking (SDN)~\cite{sharing-6}. A controller is placed on a Packet Delivery Network (PDN) gateway that is used to enable cooperation between mobile devices connected to the controller. The controller aims at reducing traffic on the gateway and ensuring fairness in energy consumption between mobile devices. To deal with dynamically arriving tasks, an Online Task Scheduling (OTS) algorithm was developed.

Centralized techniques are fairly common in the literature since they are easier to implement. However, they suffer from scalability and single point failures as is common in most centralized systems.

\textit{b. Distributed Control}: In the area of distributed control among edge devices, there seems to be relatively limited research. A game theoretic approach was employed as a decentralized approach for achieving the Nash equilibrium among cooperative devices~\cite{sharing-7}. The concept of the Nash equilibrium in the sharing model is taken further to develope the Multi-item Auction (MIA) model and Congestion Game (COG)-based sharing~\cite{sharing-8}.

\textit{ii. Based on Optimization}: There are different objectives in a system that employs a sharing model. For example, the sharing model at the edge can be employed in a battlefield scenario~\cite{sharing-9}. In this context, latencies need to be minimum, and the energy consumption of the devices needs to be at optimum. Based on existing research, the following techniques are considered for optimization:

\textit{a. Connectivity-aware}: The sharing model needs to know the connectivity between devices, for example, in the above battlefield scenario. A mobile device augments its computing when peer devices come within its communication range~\cite{sharing-9}. Then a probabilistic model predicts whether a task potentially scheduled on a peer device can complete execution in time when it is in the coverage of the device. Connectivity-aware techniques can be single hop, multi-hop, or opportunistic~\cite{sharing-10}.

\textit{Single Hop Techniques}: In this technique, a device receives a list of its neighbors that form a fully connected network. When a workload is shared by a device, the workload will be distributed to other devices that are directly connected to the device.

\textit{Multi-hop Techniques}: Each device computes the shortest path to every other node in the network that can share its workload. The work is usually shared with devices that may reduce the overall energy footprint. The benefit of a multi-hop technique in the sharing model compared to single hop techniques is that a larger pool of resources can be tapped into for more computationally intensive workloads. A task distribution approach using a greedy algorithm to reduce the overall execution time of a distributed workload was recently proposed~\cite{sharing-11}.

\textit{Opportunistic Techniques}: The device that needs to share its workload in these techniques checks whether its peers can execute a task when it is within the communication range. This is predicted via contextual profiling or historical data of how long a device was within the communication range of its peers. In recent research, a connectivity-aware opportunistic approach was designed such that: (i) data and code for the job can be delivered in a timely manner, (ii) sequential jobs are executed on the same device so that intermediate data does not have to be sent across the network, and (iii) there is distributed control, and jobs are loosely coupled~\cite{sharing-12}. The jobs are represented as a Directed Acyclic Graph (DAG), and the smallest component of a job is called a PNP-block that is used as the unit scheduled onto a device. In the context of Internet-of-Things (IoT) for data-centric services, it is proposed that a collection of mobile devices forms a mobile cloud via opportunistic networking to service the requests of multiple IoT sensors~\cite{sharing-13}.

\textit{b. Heterogeneity-aware}: Edge devices in a mobile cloud are heterogeneous at all levels. Therefore, the processor architecture, operating system, and workload deployment pose several challenges in facilitating cooperation~\cite{sharing-14}. There is recent research tackling heterogeneity-related issues in mobile networks. For example, a work sharing approach named Honeybee was proposed in which cycles of heterogeneous mobile devices are used to serve the workload from a given device~\cite{sharing-15}. The approach accounts for devices leaving/joining the system. Similarly, a framework based on service-oriented utility functions was proposed for managing heterogeneous resources that share tasks~\cite{sharing-16}. A resource coordinator delegates tasks to resources in the network so that parameters, such as gain and energy, are optimized using convex optimization techniques.

\textit{c. Security-aware}: A technique to identify and isolate malicious attacks that could exist in a device used in the sharing model, referred to as HoneyBot, has been proposed~\cite{sharing-17}. A few of the devices in a mobile network are chosen as HoneyBots for monitoring malicious behavior. In the provided experimental results, a malicious device can be identified in 20 minutes. Once a device is identified to be malicious, it is isolated from the network to keep the network safe.

d. Fairness-aware: Fairness has been defined as a multi-objective optimization problem. The objectives are to reduce the drain on the battery of mobile devices so as to prolong the network lifetime, and at the same time improve the performance gain of the workload shared between devices~\cite{sharing-18}. The processing chain of mobile applications was modeled as a DAG and assumed that each node of the DAG is an embarrassingly parallel task. Each task was considered as a Multi-objective Combinatorial Bottleneck Problem (M-CBP) solved using a heuristic technique.

\textit{iii. Based on Cooperation}: Edge devices can share workloads (a) either in a less defined environment that is based on ad hoc cooperation, or (b) in a more tightly coupled environment where there is infrastructure to facilitate cooperation.

\textit{a. Ad Hoc Cooperation}: Setting up ad hoc networks for device-to-device communication is not a new area of research. Ad hoc cooperation has been reported for MCC in the context of the sharing model for the edge~\cite{sharing-19}. There is recent research that has coined the term “transient clouds,” in which neighboring mobile devices form an ad hoc cloud and the underlying task management algorithm is based on a variant of the Hungarian method~\cite{sharing-20}.

\textit{b. Infrastructure-based Cooperation}: There is research on the federation of devices at the edge of the network to facilitate cooperation~\cite{sharing-21}. This results in more tightly coupled coalitions than ad hoc clouds, and more cost effectiveness than dedicated micro cloud deployment.

\subsubsection{\textbf{Offloading}}
\label{sec:offloading}
\begin{figure*}
	\centering
	\includegraphics[width=0.95\textwidth]{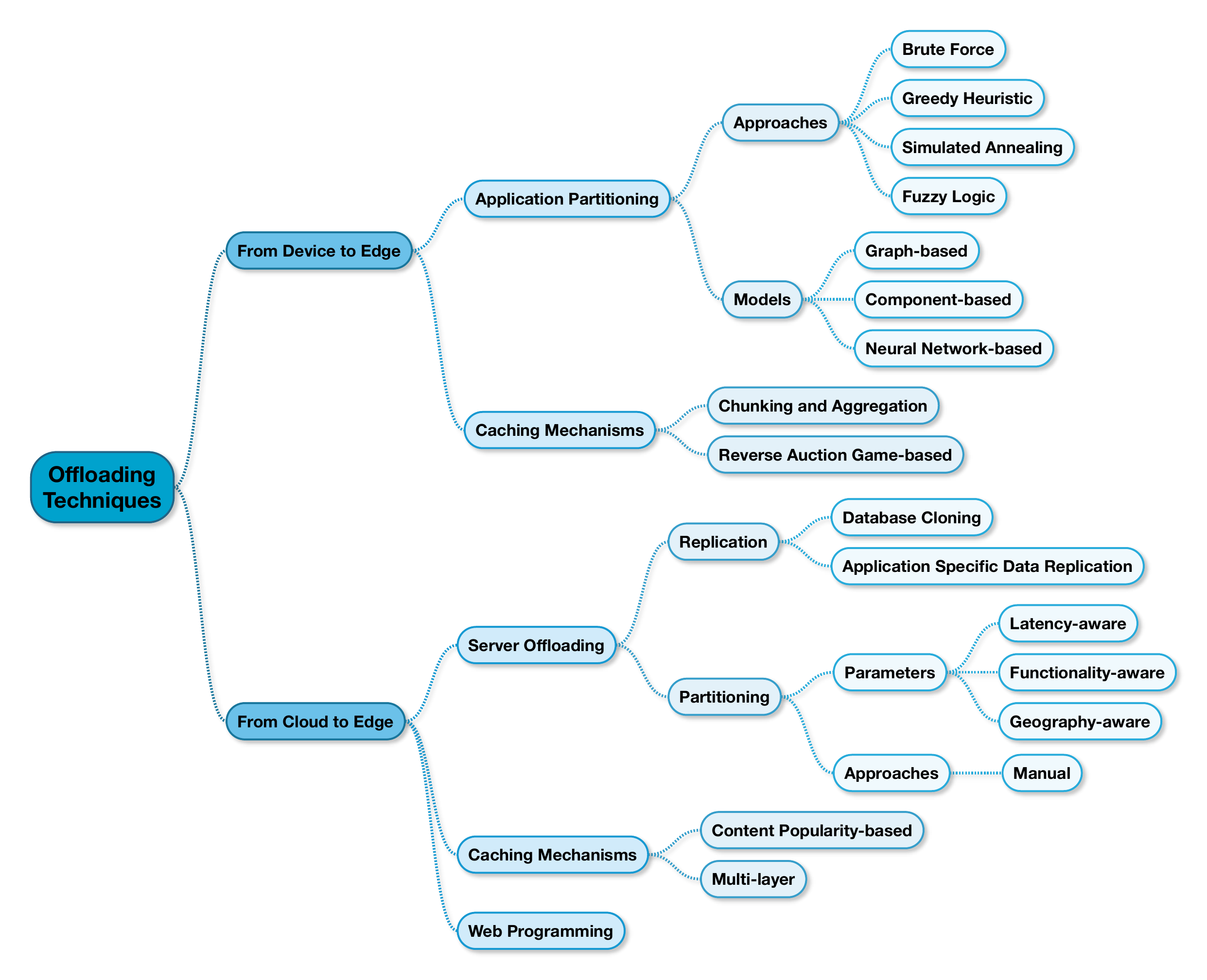}
	\caption{A classification of offloading techniques}
	\label{fig:offloading}
\end{figure*}

Offloading is a technique in which a server, an application, and the associated data are moved on to the edge of the network. This either augments the computing requirements of individual or a collection of user devices, or brings services in the cloud that process requests from devices closer to the source. Research in offloading can be differentiated in the following two ways, as presented in Figure~\ref{fig:offloading}:

\textit{i. Offloading from User Device to Edge}: This technique augments computing in user devices by making use of edge nodes (usually a single hop away). The two main techniques used are application partitioning and caching mechanisms.

\textit{a. Application Partitioning}: One example of offloading from devices to the edge via application partitioning is in the GigaSight architecture in which Cloudlet VMs \cite{satyanarayanan2009case} are used to process videos streamed from multiple mobile devices~\cite{off-1}. The Cloudlet VM is used for denaturing, a process of removing user-specific content for preserving privacy. The architecture employed is presented as a Content Delivering Network (CDN) in reverse. In this survey, we discuss the following four approaches and three models used for application partitioning.

\textit{Approaches}: Four approaches are considered, namely, brute force, greedy heuristic, simulated annealing, and fuzzy logic.

Brute Force: There is a study under the umbrella of ENGINE that proposes an exhaustive brute force approach, in which all possible combinations of offloading plans (taking the cloud, edge nodes, and user devices) are explored~\cite{off-2}. The plan with the minimum execution time for a task is then chosen. This approach simply is not a practical solution given the time needed to derive a plan, but instead could provide insight into the search space.

Greedy Heuristic: ENGINE also incorporates a greedy approach that focuses on merely minimizing the time taken for completing the execution of a task on the mobile device~\cite{off-2}. An offloading plan is initially generated for each task on a mobile device, and then iteratively refined to keep the total monetary costs low. Similarly, FogTorch, a prototype implementation of an offloading framework, uses a greedy heuristic approach for deriving offloading plans~\cite{off-3}.

Simulated Annealing: Another approach is simulated annealing, in which the search space is based on the utilization of fog and cloud nodes, total costs, and the completion time of an application to obtain an offloading plan that minimizes the costs and the completion time of the task~\cite{off-2}.

Fuzzy Logic: There is research highlighting that an application from a user device can be partitioned and placed on fog nodes using fuzzy logic~\cite{off-3a}. The goal is to improve the Quality-of-Experience (QoE) measured by multiple parameters such as service access rate, resource requirements, and sensitivity toward data processing delay. Fuzzy logic is used to prioritize each application placement request by considering the computational capabilities of the node.

\textit{Models}: The three underlying models used for application partitioning from devices to the edge are graph-based, component-based, and neural network-based.

Graph-based: CloneCloud employs a graph-based model for the automated partitioning of an application~\cite{off-4}. Applications running on a mobile device are partitioned and then offloaded onto device clones in the cloud. In the run-time, this concept translates to migrating the application thread onto the clone, after which it is brought back onto the original mobile device. Similarly, in another graph-based approach, each mobile application task to be partitioned is represented as a Directed Acyclic Graph (DAG)~\cite{off-5}. The model assumes that the execution time, migration time, and data that need to be migrated for each task are known a priori via profiling. Aspect-oriented programming is then used to obtain traces of sample benchmarks. Thereafter, a trace simulation is used to determine whether offloading to the edge nodes would reduce execution time.

Component-based: In this case, the functionalities of an application (a web browser) that runs on a device are modeled as components that are partitioned between the edge server and the device~\cite{off-6}. The example demonstrated is Edge Accelerated Web Browsing (EAB), in which individual components of a browser are partitioned across the edge and the device. The contents of a web page are fetched and evaluated on the edge while the EAB client merely displays the output.

Neural Network-based: Recent research highlights the distribution of deep neural networks across user devices, edge nodes, and the cloud ~\cite{off-7, off-8}. The obvious benefit is that the latency of inferring from a deep neural network is reduced for latency-critical applications without the need to transmit images/video far from the source. Deep networks typically have multiple layers that can be distributed over different nodes. The Neurosurgeon framework models the partitioning between layers that will be latency- and energy-efficient from end-to-end~\cite{off-7}. The framework predicts the energy consumption at different points of partitioning in the network and chooses a partition that minimizes data transfer and consumes the least energy. This research was extended towards distributing neural networks across geographically distributed edge nodes~\cite{off-8}.

\textit{b. Caching Mechanisms}: This is an alternative to application offloading. In this mechanism, a global cache is made available on an edge node that acts as a shared memory for multiple devices that need to interact. This survey identifies two such mechanisms, namely chunking and aggregation, and a reverse auction game-based mechanism.

\textit{Chunking and Aggregation}: The multi Radio Access Technology (multi-RAT) was proposed as an architecture for upload caching. In this model, VMs are located at the edge of the network, and a user device uploads chunks of a large file onto them in parallel ~\cite{off-9}. Thereafter, an Aggregation VM combines these chunks that are then moved onto a cloud server.

\textit{Reverse Auction Game-based}: An alternate caching mechanism based on cooperation of edge nodes was proposed in~\cite{off-10}. The users generate videos that are shared between the users via edge caching. The mechanism uses a reverse auction game to incentivize caching.

\textit{ii. Offloading from the Cloud to the Edge}: The direction of data flow is opposite that considered above; in this case, a workload is moved from the cloud to the edge. There are three techniques that are identified in this survey including server offloading, caching mechanisms, and web programming.

\textit{a. Server Offloading}: A server that executes on the cloud is offloaded to the edge via either replication or partitioning. The former is a naive approach that assumes that a server on the cloud can be replicated on the edge.

\textit{Replication}: Database cloning and application data replication are considered.

Database Cloning: The database of an application may be replicated at the edge of the network and can be shared by different applications or users~\cite{off-11}.

Application-specific Data Replication: In contrast to database cloning, a specific application may choose to bring data relevant to the users to the edge for the seamless execution of the application~\cite{off-12}. However, both database cloning and application-specific data replication  assume that edge nodes are not storage-limited, so they may not be feasible in resource-constrained edge environments.

\textit{Partitioning}: We now consider the server partitioning parameters that are taken into account in offloading from the cloud to the edge. The parameters considered in partitioning are functionality-aware, geography-aware, and latency-aware.

Functionality-aware: Cognitive assistance applications, for example Google Glass, are latency-critical applications, and the processing required for these applications cannot be provided by the cloud alone. Therefore, there is research on offloading the required computation onto Cloudlet VMs to meet the processing and latency demands of cognitive assistance applications~\cite{off-13}. The Gabriel platform built on OpenStack++ is employed for VM management via a control VM, and for deploying image/face recognition functionalities using a cognitive VM on Cloudlet.

Geography-aware: The service requests of online games, such as PokeMon Go, are typically transmitted from user devices to a cloud server. Instead of sending traffic to data centers, the ENORM framework partitions the game server and deploys it on an edge node~\cite{off-14}. Geographical data relevant to a specific location is then made available on an edge node. Users from the relevant geographical region connect to the edge node and are serviced as if they were connected to the data center. ENORM proposes an auto-scaling mechanism to manage the workload for maximizing the performance of containers that are hosted on the edge by periodically monitoring resource utilization.

Latency-aware: Similar to ENORM, a study by Báguena et al. aimed at partitioning the back-end of an application logic traditionally located on clouds so as to service application requests in real-time~\cite{off-19}. In the proposed hybrid edge-assisted execution model for LTE networks, application requests are serviced by both the cloud and the edge networks based on latency requirements. This differs from the ENORM framework, in which the server is partitioned along geographical requirements.


b. Caching Mechanisms: Content popularity and multi-layer caching are identified.

\textit{Content Popularity-based}: Content-Delivery Networks (CDNs) and ISP-based caching are techniques employed to alleviate congestion in the network when downloading apps on user devices. However, there are significant challenges arising from the growing number of devices and apps. A study by Bhardwaj et al. presented the concept of caching mechanisms specific to apps on edge nodes, such as routers and small cells, referred to as eBoxes~\cite{off-16}. This concept is called AppSachets and employs two caching strategies: based on popularity and based on the cost of caching. The research was validated on Internet traffic originating from all users at the Georgia Institute of Technology for a period of 3 months.

Similarly, there is research aimed at caching data at base stations that will be employed in 5G networks~\cite{off-17}. To achieve this, traffic is monitored to estimate content popularity using a Hadoop cluster. Based on the estimate, content is proactively cached at a base station.

\textit{Multi-layer Caching}: Multi-layer caching is a technique used in content delivery for Wireless Sensor Networks (WSNs)~\cite{off-18}. The model assumes that a global cache is available at a base station that can cache data from data centers, and that localized caches are available on edge nodes. Two strategies are employed in this technique. The first is uncoded caching, in which each node is oblivious of the cache content of other nodes, and therefore no coordination of data is required. The second technique is coded caching, in which the cached content is coded such that all edge nodes are required to encode the content for the users.

c. Web Programming: Traditional web programming makes use of the client-server model, but there is research highlighting the use of web programming that makes use of the client-edge-server architecture. The Spaceify ecosystem enables the execution of Spacelets on edge nodes that are embedded JavaScripts that use the edge nodes to execute tasks to service user requests~\cite{off-15}. An indoor navigation use-case is demonstrated for validating the Spaceify concept.


\subsection{Control}
\begin{figure}
	\centering
	\includegraphics[width=0.5\textwidth]{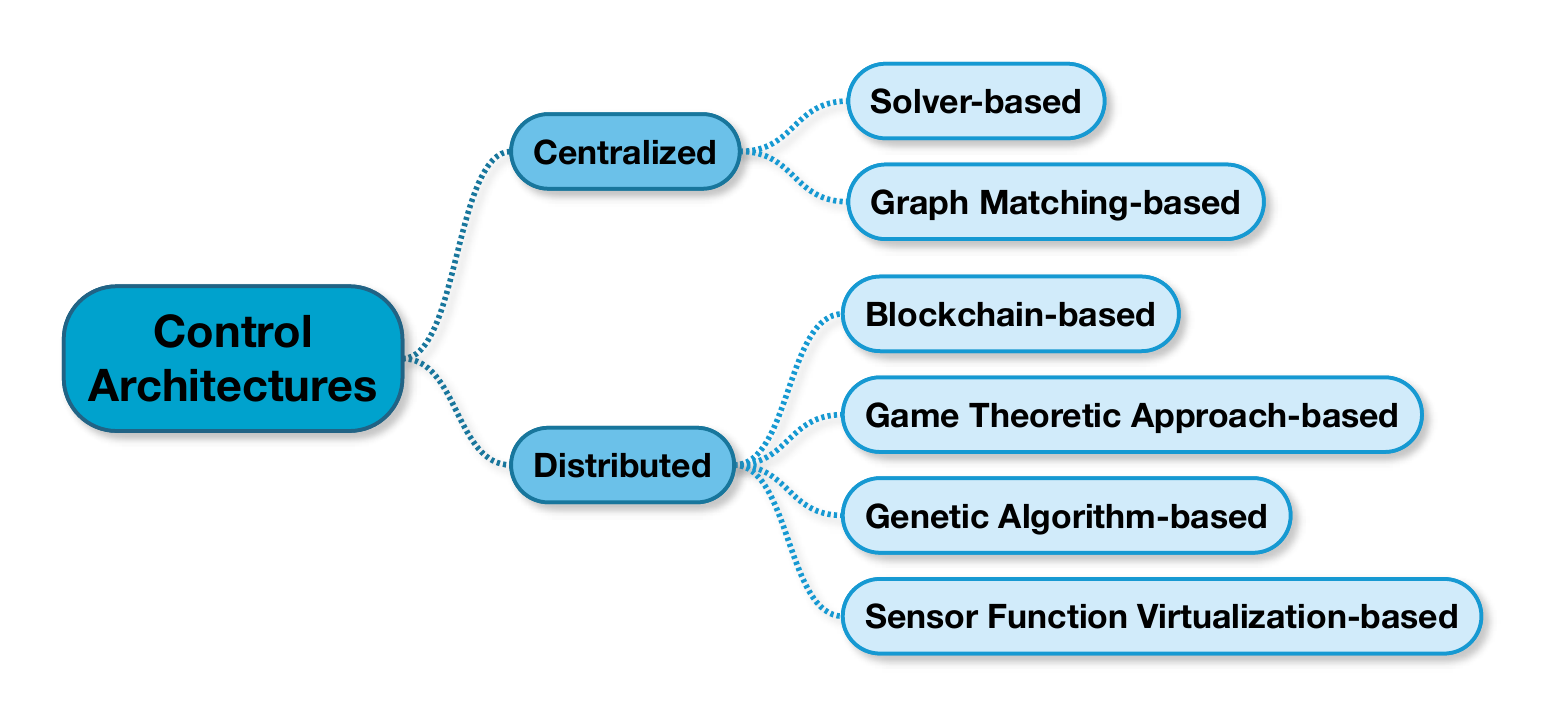}
	\caption{A classification of control architectures for resource management in fog/edge computing}
	\label{fig:control}
\end{figure}

A second method for classifying architectures for resource management in fog/edge environments is based on control of the resources. This survey identifies two such architectures, namely centralized and distributed control architectures, as shown in Figure~\ref{fig:control}. Centralized control refers to the use of a single controller that makes decisions on the computations, networks, or communication of the edge resources. On the contrary, when decision-making is distributed across the edge nodes, we refer to the architecture as distributed. This section extends the discussion on control techniques that was previously presented on sharing techniques in the survey.

\subsubsection{\textbf{Centralized}}
\label{sec:centralized}
There is a lot of research on centralized architectures, but we identify two centralized architectures, namely, (i) solver-based, and (ii) graph matching-based.

\textit{i. Solver-based}: Mathematical solvers are commonly used for generating deployment and redeployment plans for scheduling workloads in grids, clusters, and clouds. Similar approaches have been adopted for edge environments. For example, a Least Processing Cost First (LPCF) method was proposed for managing task allocation in edge nodes~\cite{control-1}. The method is underpinned by a solver aimed at minimizing processing costs and optimizing network costs. The solver is executed on a centralized controller for generating the assignment plan.

\textit{ii. Graph Matching-based}: An offloading framework that accounts for device-to-device and cloud offloading techniques was proposed ~\cite{control-2}. Tasks were offloaded via a three-layer graph-matching algorithm that is first constructed by taking the offloading space (mobiles, edge nodes, and the cloud) into account. The problem of minimizing the execution time of the entire task is mapped onto the minimum weight-matching problem in the three-layer graph. A centralized technique using the Blossom algorithm was used to generate a plan for offloading.

\subsubsection{\textbf{Distributed}}
\label{sec:distributed}
Four distributed architectures are identified: (i) blockchain-based, (ii) game theoretic-based, (iii) genetic algorithm-based, and (iv) sensor function virtualization-based.

\textit{i. Blockchain-based}: Blockchain technology is used as an underpinning technique for implementing distributed control in edge computing systems~\cite{control-3}. The technique is built on the IEC 61499 standard that is a generic standard for distributed control systems. In this model, Function Blocks, an abstraction of the process, was used as an atomic unit of execution. Blockchains make it possible to create a distributed peer-to-peer network without having intermediaries, and therefore naturally lend themselves to the edge computing model in which nodes at the edge of the network can communicate without mediators. The Hyperledger Fabric, a distributed ledger platform used for running and enforcing user-defined smart contracts securely, was used.

\textit{ii.	Game Theoretic Approach-based}: The game theoretic approach is used for achieving distributed control for offloading tasks in the multi-channel wireless interference environment of mobile-edge cloud computing~\cite{control-4}. It was demonstrated that finding an optimal solution via centralized methods is NP-hard. Therefore, the game theoretic approach is very suitable in such environments. The Nash equilibrium was achieved for distributed offloading, while two metrics, namely the number of benefitting cloud users and the system-wide computational overhead, were explored to validate the feasibility of the game theoretic approach over centralized methods.

\textit{iii. Genetic Algorithm-based}: Typically, in IoT-based systems, the end devices are sensors that send data over a network to a computing node that makes all the decision regarding all aspects of networking, communication, and computation. The Edge Mesh approach aims at distributing decision-making across different edge nodes~\cite{control-5}. For this purpose, Edge Mesh uses a computation overlay network along with a genetic algorithm to map a task graph onto the communication network to minimize energy consumption. The variables considered in the genetic algorithm are the Generation Gap used for crossover operations, mutation rate, and population size.

\textit{iv.	Sensor Function Virtualization-based}: Sensor Function Virtualization (SFV) is a visionary concept in which decision-making can be modularized and deployed anywhere in an IoT network~\cite{control-6}. The advantage of the SFV technique is that modules can be added at runtime on multiple nodes. SFV as a concept is still in infancy and needs to be demonstrated in a real world IoT testbed.


\subsection{Tenancy}
A third method for classifying architectures for resource management in fog/edge environments is tenancy. The term tenancy in distributed systems refers to whether or not underlying hardware resources are shared between multiple entities for optimizing resource utilization and energy efficiency. A single-tenant system refers to the exclusive use of the hardware by an entity. Conversely, a multi-tenant system refers to multiple entities sharing the same resource. An ideal distributed system that is publicly accessible needs to be multi-tenant.

The OpenFog reference architecture highlights multi-tenancy as an essential feature in fog/edge computing~\cite{openfog-1}. An application server may be offloaded from the cloud to the edge and service users. Therefore, the entities that share the hardware resources in this context are the applications that are hosted on the edge, and the users that are serviced by the edge server.

\begin{figure}
	\centering
	\includegraphics[width=0.48\textwidth]{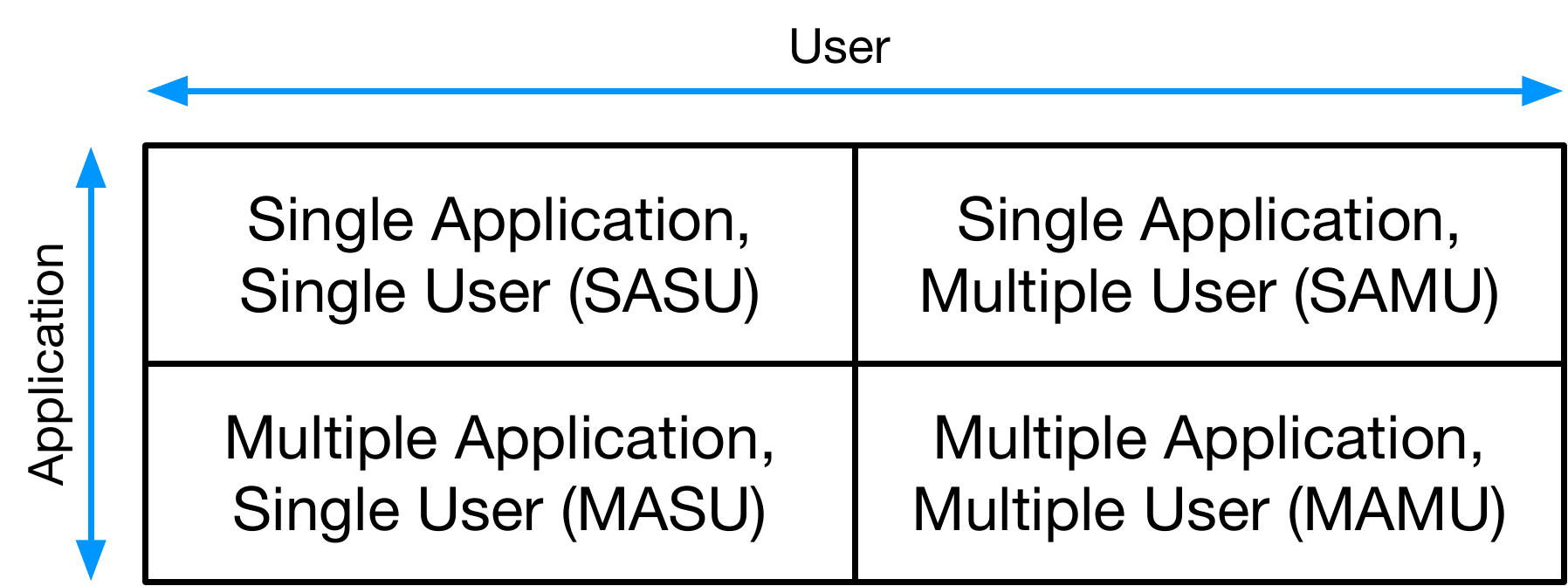}
	\caption{A taxonomy of tenancy-based architectures for resource management in Fog/Edge computing.}
	\label{fig:tenancy}
\end{figure}

In this article, we propose a classification of tenancy in fog/edge computing in two dimensions - applications and users. As shown in Figure~\ref{fig:tenancy}, the followings are the four possibilities in the taxonomy:

\begin{enumerate}[leftmargin=0.4cm]
\item [i.] \textit{Single Application, Single User (SASU)}: The edge node executes one application, and only the user can connect to the application. The application and the user solely use the hardware resources. The infrastructure is likely to be a private experimental test-bed.
\item [ii.] \textit{Single Application, Multiple User (SAMU)}: The edge node executes one application that supports multiple users. Although the underlying hardware resources are not shared among applications, there is a higher degree of sharing than SASU since multiple user requests are serviced by the edge node.
\item [iii.] \textit{Multiple Application, Single User (MASU)}: The edge node hosts multiple applications, but each application can only support a single user. This form of tenancy may be used for experimental purposes (or stress-testing the system) during the development of an ideal infrastructure.
\item [iv.] \textit{Multiple Application, Multiple User (MAMU)}: The edge node hosts multiple applications, and many users can connect to an individual application. This is an ideal infrastructure and is representative of a publicly accessible infrastructure.
\end{enumerate}

There are two techniques that support multi-tenancy, namely, system virtualization and network slicing.

1) \textit{System Virtualization}: At the system level, virtualization is a technique employed to support multi-tenancy. A variety of virtualization technologies are currently available such as traditional virtual machines (VMs) and containers (considered in Section \ref{System Virtualization}). VMs have a larger resource footprint than containers. Therefore, lightweight virtualization currently utilized in edge computing incorporates the latter~\cite{off-14, virtualization-1, virtualization-2}. Virtualization makes it possible to isolate resources for individual applications, whereby users can access applications hosted in a virtualized environment. For example, different containers of multiple applications may be concurrently hosted on an edge node.

2) \textit{Network Slicing}: At the network level, multiple logical networks can be run on top of the physical network, so that different entities with different latency and throughput requirements may communicate across the same physical network~\cite{slicing-1}. The key principles of Software Defined Networking (SDN) and Network Functions Virtualization (NFV) form the basis of slicing (considered in Section \ref{Network Virtualization}). The ongoing European project SESAME\footnote{ \url{http://www.sesame-h2020-5g-ppp.eu/Home.aspx}} (Small cells coordination for Multi-tenancy and Edge services) tackles the challenges posed by network slicing. The network bandwidth may also be partitioned across tenants, and also referred to as slicing. EyeQ is a framework that supports fine-grained control of network bandwidth for edge-based applications~\cite{slicing-2}. The framework provides end-to-end minimum bandwidth guarantees, thereby providing an efficient implementation for network performance isolation at the edge.


\section{Infrastructure}
\label{sec:infrastructure}
The infrastructure for fog/edge computing provides facilities comprising hardware and software to manage the computation, network, and storage resources \cite{confais2017object} for applications utilizing the fog/edge. In this article, the infrastructure for resource management in fog/edge computing is classified into the following three categories:

\begin{itemize}[leftmargin=0.3cm]
\item \textit{Hardware}: Recent studies in fog/edge computing suggest exploiting small-form-factor devices such as network gateways, WiFi Access Points (APs), set-top boxes, cars, and even drones as compute servers for resource efficiency \cite{stojmenovic2014Fog}. Recently, these devices are being equipped with single-board computers (SBCs) that offer considerable computing capabilities. Fog/edge computing also utilizes commodity products such as desktops, laptops, and smartphones.

\item \textit{System software}: System software runs directly on fog/edge hardware resources such as the CPU, memory, and network devices. It manages resources and distributes them to the fog/edge applications. Examples of system software include operating systems and virtualization software.

\item \textit{Middleware}: Middleware runs on an operating system and provides complementary services that are not supported by the system software. The middleware coordinates distributed compute nodes and performs deployment of virtual machines or containers to each fog/edge node.

\end{itemize}

\begin{figure}
	\centering
	\includegraphics[width=0.48\textwidth]{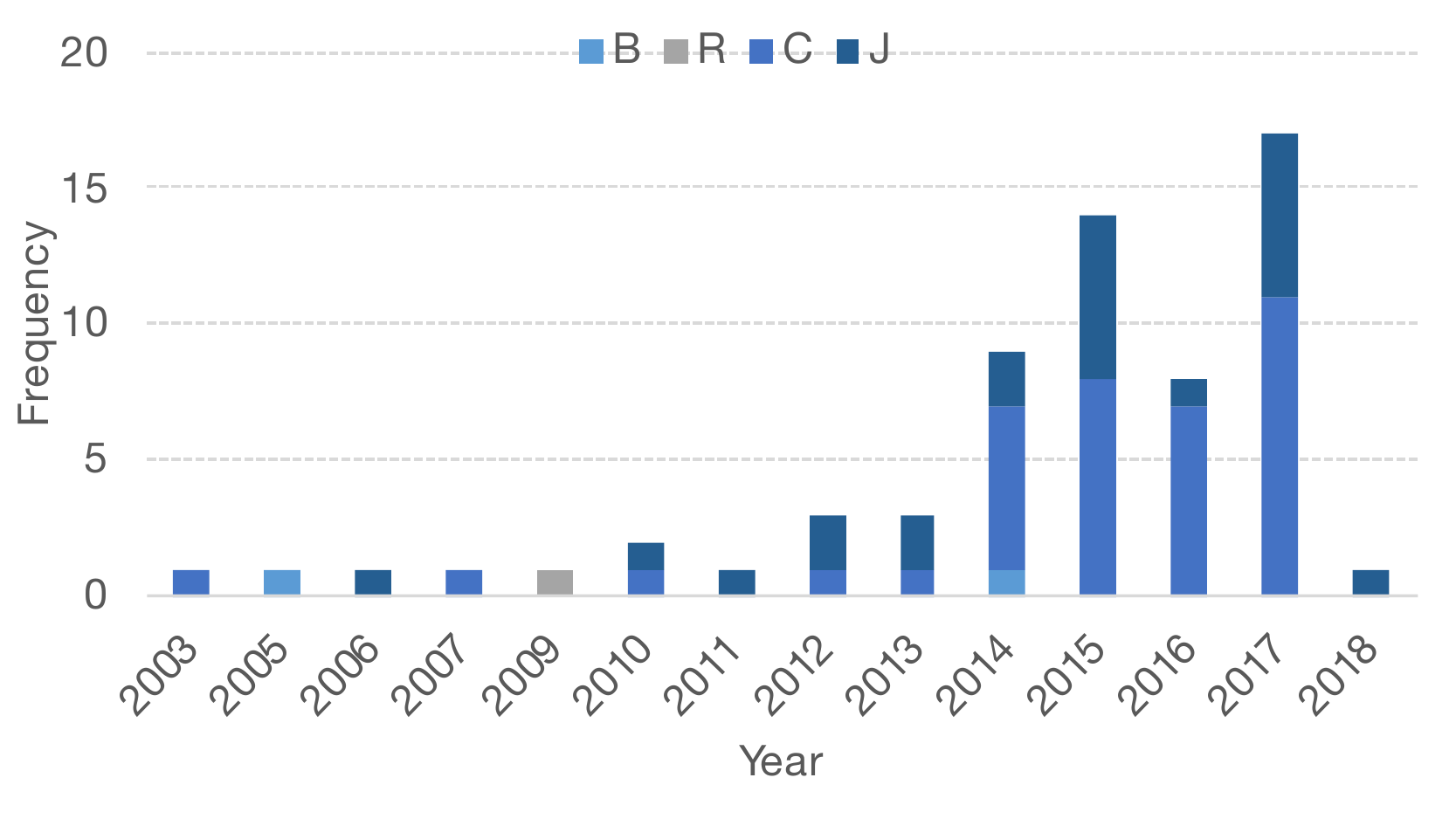}
	\caption{A histogram of publications reviewed for the classification of the infrastructure for resource management in fog/edge computing. Legend: B - books or book chapters; R - reports, including articles available on pre-print servers or white papers; C - conference or workshop papers; J - journal or magazine articles.}
	\label{fig:histogram-infrastructure}
\end{figure}

This section reviewed 63 research publications to obtain the classification of the infrastructure shown in the histogram in Figure~\ref{fig:histogram-infrastructure}. 83\% of publications were published since 2013.

\subsection{Hardware}
\label{Hardware}
\begin{figure}
	\centering
	\includegraphics[width=0.5\textwidth]{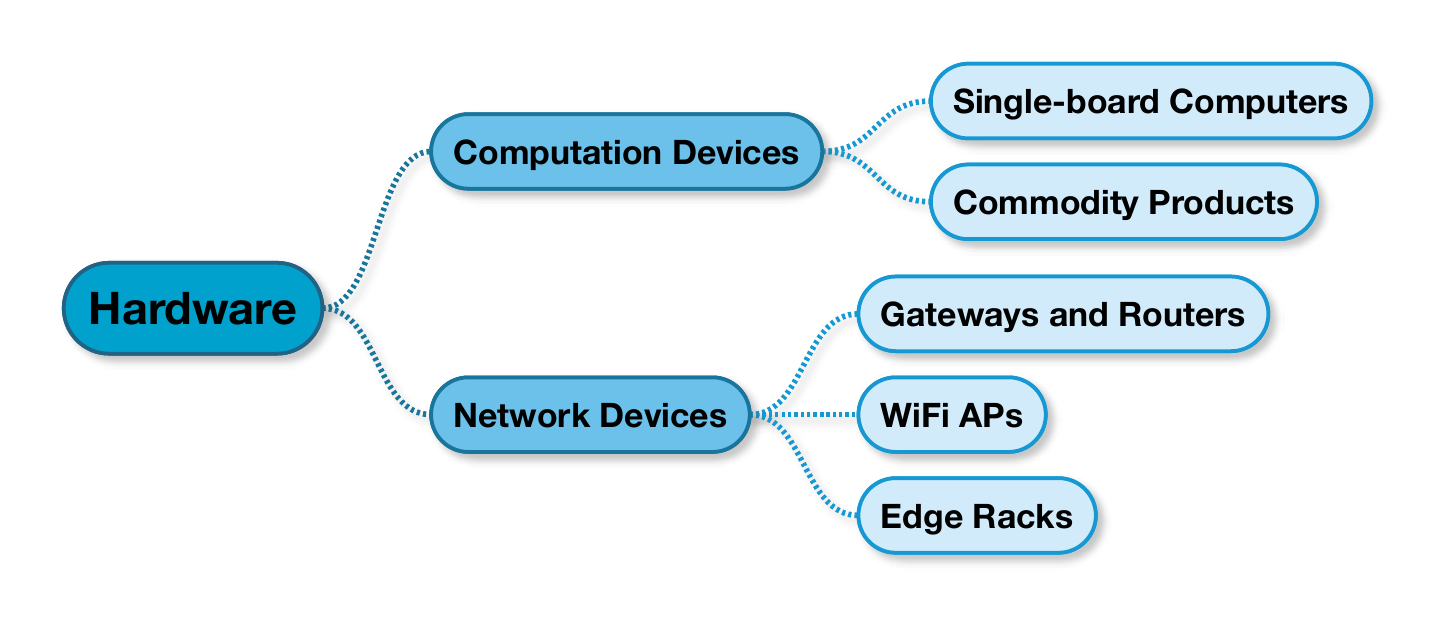}
	\caption{A classification of hardware}
	\label{fig:hardware}
\end{figure}

Fog/edge computing forms a computing environment that uses low-power mobile devices, home gateways, and routers. These small-form-factor devices nowadays have competent computing capabilities and are connected to the network. The combination of these small compute servers enables a cloud computing environment that can be leveraged by a rich set of applications processing Internet of Things (IoT) and cyber-physical systems (CPS) data. Hardware used for fog/edge computing can be classified in two ways as shown in Figure~\ref{fig:hardware}.

\subsubsection{Computation Devices}
\label{Computation Devices}
Computation devices for the fog/edge include single-board computers and commodity products that are designed for processing fog/edge data.

\textit{i. Single-board Computers}: Single-board computers (SBC) such as Raspberry Pi are often used as fog/edge nodes \cite{Cox2013, amento2016focusstack, bellavista2017feasibility}. An SBC is a small computer based on a single circuit board integrating a CPU, memory, network, and storage devices, and other components together. The small computer does not have expansion slots for peripheral devices. FocusStack \cite{amento2016focusstack} uses multiple Raspberry Pi boards installed in connected vehicles and drones to build a cloud system. FocusStack deploys a video sharing application where cameras in cars and drones capture moving scenes, and the Raspberry Pi boards process and share them. Bellavista et al. \cite{bellavista2017feasibility} used Raspberry Pi for IoT gateways that are close to sensors and actuators and therefore enable efficient data aggregation. Hong et al. \cite{hong2017Cloud} utilized Raspberry Pi for crowd-sourced fog computing and programmable IoT analytics.

\textit{ii. Commodity Products}: Commodity products such as desktops, laptops, and smartphones have been utilized as fog/edge nodes as well. For example, a recent study \cite{hong2016animation} attempted to build a cloud computing environment with laptops and smartphones used in classrooms, movie theaters, and cafes. As the owners of these devices do not always fully utilize the computational resources, fog computing providers may purchase the devices for reselling idle resources to other users. Hong et al. \cite{hong2016animation} developed an animation rendering service using under-utilized laptops in fog computing that offers cost-effectiveness compared to services in traditional cloud computing.

\subsubsection{Network Devices}
\label{Network Devices}

Network devices for fog/edge computing consist of gateways, routers, WiFi APs, and edge racks that are located in the edge and mainly process network traffics.

\textit{i. Gateways and Routers}: Network gateways and routers are potential devices for edge computing because they establish a data path between end users and network providers. Aazam et al. adopted a common gateway to decide whether the received data from IoT devices would be sent to data center clouds~\cite{aazam2014Fog}. Such smart gateways help in better utilization of network bandwidth.

\textit{ii. WiFi APs}: ParaDrop \cite{virtualization-1}, an edge computing framework, exploits the fact that WiFi APs or other wireless gateways are ubiquitous and always turned on.

\textit{iii. Edge Racks}: Global Environment for Network Innovations (GENI) packs network, computing, and storage resources into a single rack~\cite{gosain2016enabling}. GENI implements an edge computing environment by deploying GENI racks at several networked sites. These racks currently connect over 50 sites in the USA and are used as Future Internet and Distributed Cloud (FIDC) testbeds.

\subsection{System Software}
\label{System Software}

\begin{figure}
	\centering
	\includegraphics[width=0.5\textwidth]{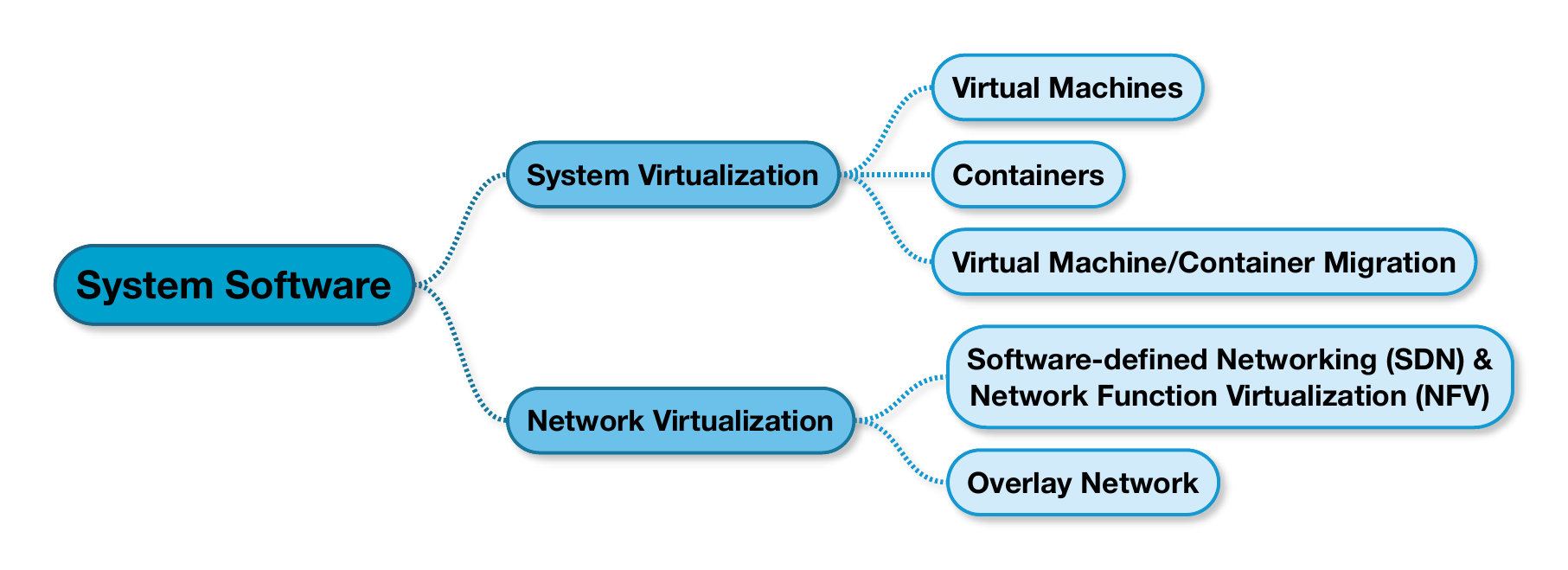}
	\caption{A classification of system software}
	\label{fig:software}
\end{figure}

System software for the fog/edge is a platform designed to operate directly on fog/edge devices and manage the computation, network, and storage resources of the devices. Examples include virtual machines (VMs) and containers. The system software needs to support multi-tenancy and isolation because fog/edge computing accommodates several applications from different tenants. System software used for fog/edge computing can be classified into two categories, system virtualization and network virtualization, as shown in Figure~\ref{fig:software}.

\subsubsection{System Virtualization}
\label{System Virtualization}

System virtualization allows multiple operating systems to run on a single physical machine. System virtualization enables fault and performance isolation between multiple tenants in the fog/edge. It partitions resources for each tenant so that one tenant cannot access other tenants' resources. The fault of a tenant, therefore, cannot affect other tenants. System virtualization also limits and accounts for the resource usage of each tenant so that a tenant cannot monopolize all the available resources in the system. This section deals with traditional virtual machines, recent containers, and VM/container migration software for supporting system virtualization.

\begin{figure*}
	\centering
	\includegraphics[width=0.85\textwidth]{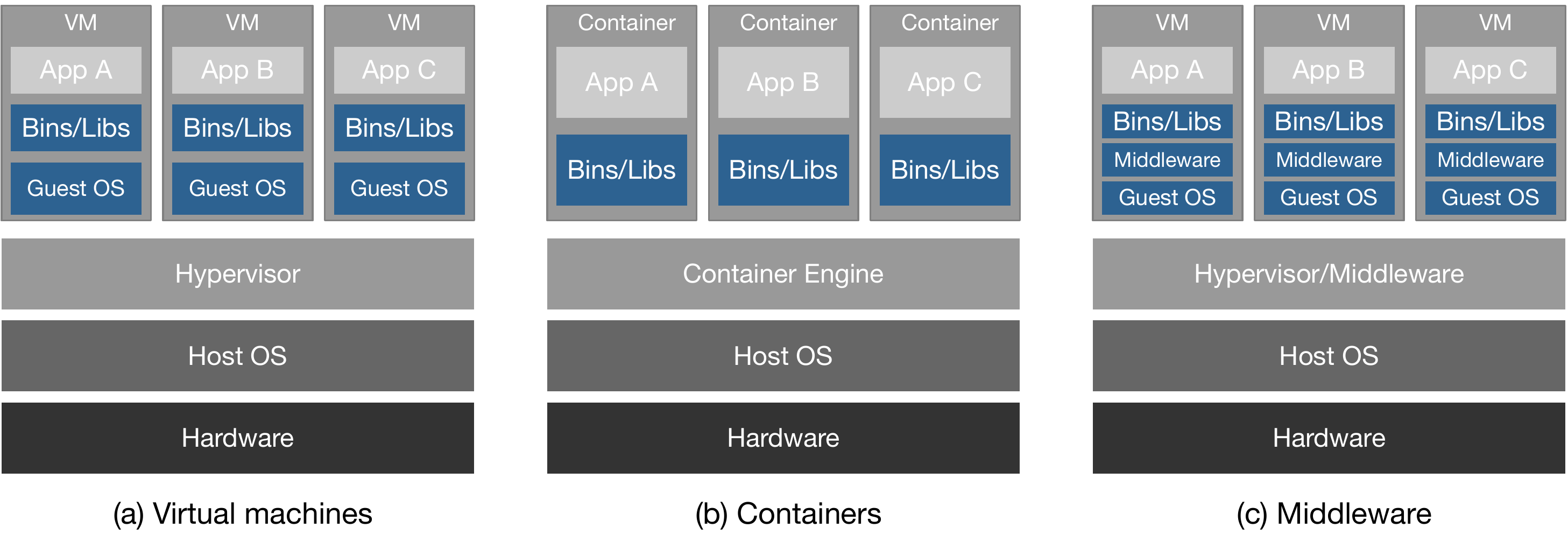}
	\caption{Architectures of virtual machines, containers, and middleware for resource management in fog/edge computing}
	\label{fig:vm_container_middleware}
\end{figure*}

\textit{i. Virtual Machines}: A Virtual Machine (VM) is a set of virtualized resources used to emulate a physical computer. Virtualized resources include CPUs, memory, network, storage devices~\cite{chiueh2005survey}, and even GPUs and FPGAs \cite{hong2017gpu}. Virtualization software called a hypervisor (e.g., Xen \cite{barham2003xen} or KVM \cite{kivity2007kvm}) virtualizes the physical resources and provides the virtualized resources in the form of a VM. The tenant installs an operating system and runs applications in the VM, regarding the VM as a real physical machine. The VM architecture is shown in the left side of Figure \ref{fig:vm_container_middleware}. Virtualization isolates the execution environment between fog/edge tenants. Each tenant maintains its own VMs, and IoT and CPS devices of the tenant send data to the VMs for processing and storage~\cite{satyanarayanan2009case, gu2017cost, wang2017elastic, gosain2016enabling}.


\emph{Cloudlet} provides an early form of fog computing by offering resource-rich VMs for mobile devices in close proximity~\cite{satyanarayanan2009case}. As a mobile device connects to a VM over a wireless LAN, Cloudlets achieve low latency when the mobile device offloads tasks to the VM.

Gu et al. \cite{gu2017cost} proposed a fog computing architecture using VMs for a medical cyber-physical system (MCPS) \cite{lee2012challenges}. To receive fast and accurate medical feedbacks, the MCPS system utilizes computational resources close to medical devices. The research utilizes low power sensors and actuators for collecting health information and then sends the collected information to a VM in the network edge (e.g., base stations) for storage and analyses. The research associates several medical devices of a tenant with a VM running in the edge.

Wang et al. \cite{wang2017elastic} implemented a real time surveillance infrastructure where surveillance cameras send images to a distributed edge cloud platform. The surveillance system launches a group of VMs to which surveillance tasks are distributed. When the load is high across the cloud, the system elastically launches new VMs to secure more computational power and network bandwidth.

GENI \cite{gosain2016enabling} provides GENI racks to realize an edge-based cloud computing platform for university campuses. Each rack consists of Layer-2 and -3 switches and compute nodes that provide VMs to university students on demand. This infrastructure is available around 50 campuses in the USA.

\textit{ii. Containers}: Containers are an emerging technology for cloud computing that provides process-level lightweight virtualization \cite{vaughan2006new, haydel2015enhancing}. Containers are multiplexed by a single Linux kernel, so that they do not require an additional virtualization layer compared to virtual machines. Although they share the same OS kernel, they still offer operating systems virtualization principles \cite{pahl2015containers} where each user is given an isolated environment for running applications. The architecture of containers is shown in the middle of Figure~\ref{fig:vm_container_middleware}.

Namespaces in Linux provide containers with their own view of the system, and \emph{cgroups} are responsible for resource management such as CPU allocation to containers. This lightweight virtualization allows containers to start and stop rapidly and to achieve performance similar to that of the native environment. In addition, containers are usually deployed with a pre-built application and its dependent libraries, focusing on Platform-as-a-Service (PaaS) that makes container-based applications easily deployed and orchestrated. Representative container tools include LXC \cite{helsley2009lxc} and Docker \cite{merkel2014docker} for building and deploying containers, and Kubernetes \cite{brewer2015kubernetes} for orchestration.

Lightweight virtualization implemented by containers facilitates the adoption of performance-limited resources in fog/edge computing nodes \cite{bellavista2017feasibility, morabito2016enabling, virtualization-1}. Bellavista et al. \cite{bellavista2017feasibility} employed Docker-based containers on a Raspberry Pi 1 board that is used as a fog node for collecting data from heterogeneous sensors in a transit vehicle or other infrastructural components. Morabito et al. \cite{morabito2016enabling} utilized single-board computers, including RaspberryPi 2, Odroid C1+, and Odroid XU4 boards as edge processing devices running Docker containers. ParaDrop \cite{virtualization-1} adopted lightweight containerization for WiFi Access Points (APs) or other wireless gateways.

In these studies, containers provide low overhead for performance-limited hardware platforms. In addition, a system administrator can package an application into a container for aggregating and processing data and send the container to any device for creating a new fog/edge node on the fly. Finally, containers allow a high density of applications due to small images, which is useful on resource-limited devices.

\textit{iii. Virtual Machine/Container Migration}:
Virtual Machine (VM) or container migration moves a running VM or container to different physical machines without affecting the applications running on the VM or container \cite{medina2014survey, ahmad2015survey, forsman2015algorithms}. For this purpose, the storage and network connectivity of the transferred VM or container also needs to be moved to the target physical machine \cite{cerroni2014live}. In fog/edge computing, location-awareness should be considered for migration performance \cite{stojmenovic2014Fog}.

INDICES \cite{shekhar2017indices} points out that server overloading needs to be addressed when a VM is migrated from a cloud data center to a fog cloud platform. INDICES considers the performance interference caused by resource contention between co-located VMs during VM migration. INDICES first identifies a user experiencing service level objective (SLO) violations and moves the user's VM to a fog cloud platform that can offer the lowest performance interference.

Bittencourt et al. \cite{bittencourt2015towards} detected the movement and behavior of a mobile device to decide where and when to migrate the user's VM among fog cloud platforms. When a user's device is disconnected from the access point of one fog cloud, the study identifies the user's location using a GPS system, and moves the user's VM to a nearby fog cloud. As data migration may incur a service suspension during migration, the research adopts a proactive technique that migrates the VM in advance, predicting the user's movement.

\subsubsection{Network Virtualization}
\label{Network Virtualization}

Network virtualization combines hardware and software network resources into a virtual network that is a software-based administrative entity for a tenant~\cite{chowdhury2010survey}.

\textit{i. Software-Defined Networking (SDN) and Network Function Virtualization (NFV)}:
A fog/edge cloud has an option to adopt software-defined networking (SDN) and network functions virtualization (NFV) for managing the network through software. SDN separates the control plane from the data plane~\cite{kreutz2015software}. The control plane decides where the traffic is sent, and the data plane forwards the traffic to the destination decided by the control plane. NFV decouples networking functions such as routing and fire-walling from the underlying proprietary hardware, and allows each of the functions to run on a VM on commodity hardware~\cite{han2015network}. NFV is a complementary concept to SDN and is independent of it, although they are often combined together in modern clouds~\cite{manzalini2013Clouds}.

A virtual network enabled by SDN and NFV interconnects fog/edge clouds that are geographically dispersed \cite{bonomi2014Fog}. The virtual network is required to support Layer 2 (L2) and Layer 3 (L3) networks, IPv4 and IPv6 protocols, and different addressing modes. Hybrid Fog and Cloud, called HFC \cite{moreno2017cross}, extends the BEACON \cite{moreno2015beacon} project that implements a federated cloud network for the efficient and automated provision of virtual networks to distributed fog/edge clouds. The framework installs an HFC agent in each cloud that manages the control plane implemented by an SDN technology. The HFC agent also implements required Virtual Network Functions (VNFs) such as virtual switches and routers in order to interconnect the distributed clouds.

Constructing SDN and NFV in fog/edge platforms implies that clients can leverage elastic virtualized environments where all VMs for the same tenant can be in the same virtual LAN (VLAN) even if they are located in different areas. Wang et al. built an urban video surveillance system that exploits Virtualized Network Functions (VNFs) VMs as computational units for video analysis algorithms~\cite{wang2017elastic}. More VMs can be allocated to higher priority tasks, and the source data can be sent between VMs by virtual switches controlled by the SDN routing strategies.

Conventional mobile clouds that offload tasks from mobile devices to centralized data centers are moving their applications to fog/edge clouds so as to reduce processing latency. NFV in fog/edge devices constructs a virtualized network infrastructure where computational resources can be scaled on the infrastructure based on demand. Yang et al. proposed a set of algorithms for dynamic resource allocation in such an NFV-enabled mobile fog/edge cloud~\cite{yang2016seamless, yang2018cost}. An offline algorithm estimates the desired response time with minimum resources, and the auto-scaling and load-balancing algorithm makes provision for workload variations. When the capacity violation detection algorithm identifies a failure of the auto-scaling mechanism, a network latency constraint greedy algorithm initializes an NFV-enabled edge node to cope with the failure.

SDN is also applied to inter-vehicle communication using fifth generation (5G) vehicular networks or Vehicular Adhoc Network (VANET)~\cite{vinel2017emerging, truong2015software}. In this context, SDN can efficiently manage connected vehicles, called a vehicular neighbor group, with efficient member selection, group establishment, and flexible resource scheduling. 5G-SDVN abstracts vehicles on a 5G network as SDN switches and simplifies network management~\cite{huang2017exploring}. In this study, mobile fog computing is also exploited by considering vehicles as mobile users. As with 5G-SDVN, FSDN VANET applies both SDN and fog computing to connected vehicles on a VANET~\cite{truong2015software}. VANET is limited; it has long delays and unbalanced flow traffic when the number of vehicles increases. The separation of the control and data planes in SDN simplifies network management as the number of vehicles increases, while fog computing improves VANET services with additional computational capabilities.

In recent fog computing use cases, data tend to be internally generated and consumed between sensors \cite{ivanov2016gravity}. In this setup, each fog node is expected to act as a wireless router to transfer data between sensors. Hakiri et al. \cite{hakiri2017managing} employed SDN for managing wireless fog networks. SDN generally adopts a centralized control plane, but the authors pointed out that this can be a single point of failure and might deteriorate reliability. This study developed a hybrid control plane in which a centralized controller manages the entire network, and additional controllers are attached during runtime to serve as backup should the centralized controller fail.

Huge traffic volumes from IoT devices can disrupt conventional IoT networks. SDN architectures can help to alleviate this problem. Xu et al. incorporated a Message Queuing Telemetry Transport (MQTT) that is an application layer protocol for IoT, with SDN-enabled fog computing~\cite{agg-12, karagiannis2015survey}. MQTT consists of publishers, subscribers, and the broker. The broker receives messages from a publisher and relays the published messages to subscribers. The study developed an SDN-based proxy broker where the broker acts as a control plane. The broker aggregates traffics from clients for effective transmission and utilizes an Open vSwitch (OVS) to forward traffic.

\textit{ii. Overlay Network}:
An overlay network is a virtual network that is based on an underlying physical network and that provides additional network services (for example, peer-to-peer networks). Nodes in the overlay network are connected by virtual links to enable a new data path over physical links.

Koala \cite{tato2017designing} proposed an overlay network for decentralized edge computing. Different from cloud computing, there is no controller in decentralized edge computing, so each node has a limited view of the network. Koala built an overlay network to encourage collaboration between the decentralized nodes. However, proactively maintaining the overlay network incurs significant network traffic for identifying nodes joining or leaving the network. This is addressed by injecting maintenance messages into general applications traffic. Frugal ~\cite{aditya2017frugal} focused on constructing an overlay network for online social networks. Frugal analyzed the social graphs between users and built a degree-constrained overlay topology using minimum degree-constrained spanning trees.

\subsection{Middleware}

\begin{figure}
	\centering
	\includegraphics[width=0.5\textwidth]{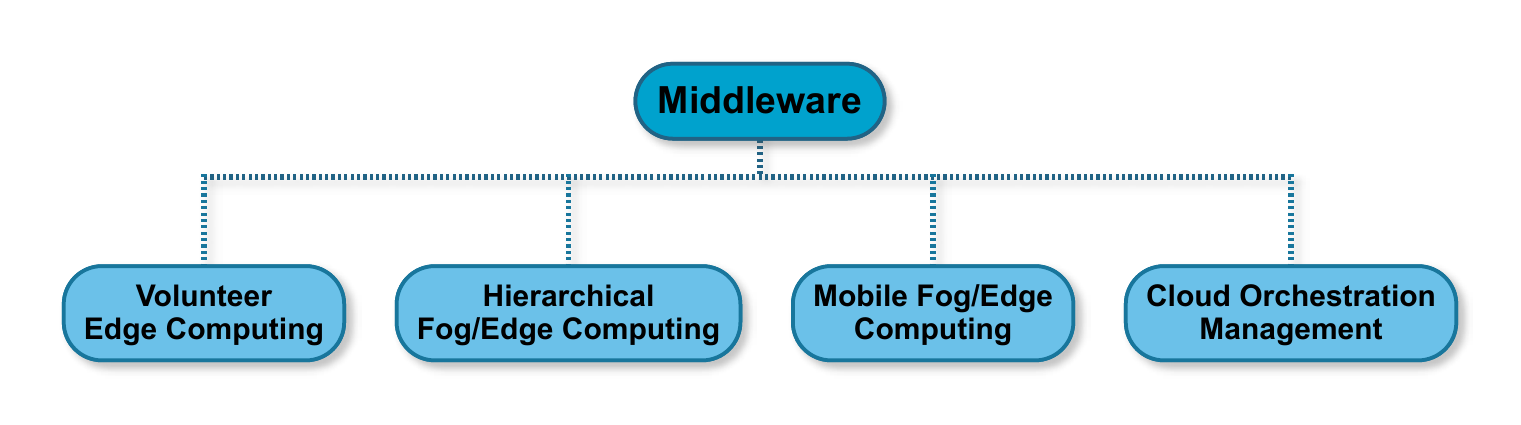}
	\caption{A classification of middleware}
	\label{fig:middleware}
\end{figure}

Middleware provides complementary services to system software. Middleware in fog/edge computing provides performance monitoring, coordination and orchestration, communication facilities, protocols, and so on. Middleware used for fog/edge computing can be classified into four categories, as shown in Figure~\ref{fig:middleware}. The architecture of middleware is shown on the right side of Figure \ref{fig:vm_container_middleware}.

\subsubsection{Volunteer Edge Computing} Nebula is middleware software that enables a decentralized edge cloud that consists of volunteer edge nodes that contribute resources~\cite{chandra2013decentralized, ryden2014nebula}. Nebula comprises four major components. First, the Nebula Central provides a web-based portal for volunteer nodes and users to join the cloud and to deploy applications. Second, the DataStore, a data storage service, enables location-aware data processing. Third, the ComputePool offers computational resources to the volunteer edge nodes. Finally, the Nebula Monitor performs computation and network performance monitoring. In Nebula, the ComputePool coordinates with the DataStore to offer compute resources that have proximity to the input data in the DataStore.

\subsubsection{Hierarchical Fog/Edge Computing}
A hierarchical fog/edge computing platform provides middleware that exploits both conventional cloud computing and recent fog/edge computing paradigms. Tasks that require prompt reaction are processed in fog/edge nodes whereas complex or long-term analysis tasks are performed at more powerful cloud nodes \cite{hong2013mobile, tang2015hierarchical, nastic2017serverless}.

\emph{Mobile Fog} facilitates hierarchical fog/edge computing. It enables easy communication between computing nodes at each hierarchical level and provides scaling capabilities during runtime~\cite{hong2013mobile}. In Mobile Fog, an application consists of three processes, each of which is mapped to a leaf node in smartphones or vehicles, an intermediate node in fog/edge computing, and a root node in the data center. Mobile Fog provides a range of APIs for communications and event handling between distributed processes. When a computing instance becomes congested, Mobile Fog creates a new instance at the same hierarchy level so as to load-balance workloads between nodes.

Tang et al. \cite{tang2015hierarchical} proposed a hierarchical fog computing platform for processing big data generated by smart cities. The platform consists of four layers. The bottom layer, Layer 4, contains a massive number of sensor nodes that are widely distributed in public infrastructures. Layer 3 consists of low-power fog/edge devices that receive raw data from the sensor nodes in Layer 4. One fog/edge device is connected to nearby sensors to provide timely data analyses. Layer 2 comprises more powerful computing nodes, each of which is connected to a group of fog/edge devices in Layer 3. Layer 2 associates temporal data with spatial data to analyze potential risky events whereas Layer 3 focuses on immediate small threats. Layer 1 is a cloud computing platform that performs long-term analyses spanning a whole city by employing Hadoop.

Nastic et al. \cite{nastic2017serverless} developed a unified edge and cloud platform for real-time data analytics. In this study, edge devices are used to execute simple data analytics such as measuring human vital signs sent from IoT mobile healthcare devices. Cloud computing receives preprocessed and filtered data from the edge devices and focuses on comprehensive data analytics to gain long-term insight about the person. The analytics function wrapper and API layer in the middleware provides a frontend for users to send and receive data to and from analytics functions in the cloud. The orchestration layer determines whether the provided data need to be processed in the edge or cloud node according to the high-level objectives of the application. Finally, the runtime mechanism layer schedules analytics functions and executes them while satisfying QoS requirements.

\subsubsection{Mobile Fog/Edge Computing}
Conventional mobile cloud computing \cite{cuervo2010maui, off-4, kosta2012thinkair} allows low-power mobile devices such as smartphones to offload their computation-intensive tasks to more powerful platforms in cloud computing. This feature can improve the user experience and save power in mobile devices. However, cloud platforms in data centers cannot support low network latency and high bandwidth. To address this limitation, developers are exploiting fog/edge computing to offload their tasks for achieving satisfactory latency and bandwidth.

FemtoClouds \cite{sharing-4} pay attention to recent powerful mobile devices such as smartphones and laptops, and form a compute cluster using these devices. A controller in FemtoClouds receives requests from users who installed the FemtoCloud service and schedules the requests in idle devices with sufficient capability. A business holder such as a coffee shop owner or a university can provide the controller. The Discovery Module in the controller discovers FemtoCloud devices and estimates the compute capacity of each one. Upon users' requests, the Execution Prediction Module predicts the completion time based on each device's execution load. The Task Assignment Module then iteratively assigns several tasks to less loaded devices to efficiently obtain the desired results.

Sensors Of Ubiquitous Life (SOUL) \cite{jang2016soul} constructs an edge-cloud for efficiently processing various sensors in mobile devices. The authors pointed out that an application in a mobile device might not know how to handle device-specific sensors. SOUL provides APIs to virtualize sensors, thereby making it possible for diverse sensors to be treated in the same way. SOUL externalizes the virtualized sensors to the edge-cloud to leverage the cloud's computational and storage services. The SOUL Engine in each mobile device manages sensor-related operations executed by the application and sends these requests to the edge-cloud. The SOUL Core in the edge-cloud performs the received requests on behalf of the device. The two entities are connected by SOUL Streams.

Silva et al. \cite{silva2017using} extended mobile cloud computing to an edge-cloud where nearby devices are connected by WiFi-Direct. The connected devices work together as a pool of computing resources for data caching and video streaming. Mobile devices that share the same interest (e.g., devices in the same sports stadium) establish a WiFi-Direct group. The cloud middleware tracks the members of the group along with their connection information and provides the content stored in each mobile device. This architecture relieves the load in the access points at a certain large venue and improves the quality of experience.

Human-driven edge computing (HEC) \cite{bellavista2018human} points out that mobile edge computing has limitations because the number of edges is not sufficient, and some highly populated areas may result in congestion on edges. To address these limitations, HEC combines mobile edge computing with mobile crowdsensing \cite{ganti2011mobile}, where smartphones or tablet computers become edge nodes, share sensor data, and analyze the data for common interest. HEC does not implement a controller, but instead exploits local one-hop communications using VM/container migration between participants. The middleware for HEC consists of two components. \emph{Elijah} is responsible for cloud resource management and  migrates a VM to an identified edge node. The \emph{Elijah} extension module additionally supports Docker-oriented containers and enables seamless VM/container migration when handoffs occur between different edge nodes.

\subsubsection{Cloud Orchestration Management}
In fog/edge computing, each device is regarded as a small compute server. The inter-device coordination for these devices is challenging compared to conventional clouds because (i) fog/edge devices have limited capabilities, (ii) the number of fog/edge devices expected to participate is greater than that of compute servers in a cloud data center, and (iii) fog/edge devices may be moving, and therefore the connectivity to the network may be intermittent~\cite{amento2016focusstack}.

While container-based cloud computing provides low overhead, a device in fog/edge computing still has resource limitations that cause the device to perform until it reaches the maximum computing capacity. The microCloud \cite{morabito2016enabling} overcomes this limitation by exploiting resources of other edge devices. It adopts the Cloudy software \cite{selimi2015Cloud}, a proprietary cloud management framework for local communities that is associated with Docker-based containers. Using the framework, a user can publish applications to a set of containers running on several edge devices. The microCloud thereby provides elasticity like other public clouds. The microCloud focused on local homogeneous devices, while Khan et al. \cite{khan2017Edge} extended the concept of the microCloud to geographically distributed and heterogeneous devices.

Edge compute nodes may consist of thousands to millions of moving devices such as cars and drones. In this scenario, it is challenging to orchestrate the management of the devices using existing cloud management platforms such as OpenStack \cite{sefraoui2012openstack}. FocusStack \cite{amento2016focusstack} introduces location-based awareness to OpenStack to deploy containers into devices that are geographically in the focus of attention. FocusStack minimizes managed devices at a single time by only paying attention to healthy devices in the target area. For this purpose, when a cloud operation specifying certain requirements is invoked by a FocusStack API, the Geocast Georouter sends broadcast messages to edge devices in the target area to ask whether the devices can satisfy the request. If the Geocast Georouter receives responses from the devices, it regards them as healthy devices currently connected to the network. The Conductor component then sends the corresponding OpenStack operation to the selected devices. The minimization of managed devices at a single time allows FocusStack to be more efficient and scalable in edge clouds.

Foggy \cite{santoro2017Foggy} provided an orchestration tool for hierarchical fog computing that consists of the cloud (the highest tier), edge Cloudlets, edge gateways, and Swarm of Things tiers (the lowest tier near sensors). The Orchestrator deploys each Application Component, which is a module of a large application in a container image, on a node in each tier that satisfies user requirements.

Studies by Vogler and Nastic et al. \cite{vogler2015leonore, nastic2014provisioning, nastic2016middleware} introduced middleware for IoT clouds. In these studies, \emph{software-defined IoT gateways (SDGs)} are defined for encapsulating infrastructure resources in a container. The IoT middleware focuses on the execution of provisioning workflows by supporting effective deployment of SDGs and customizing the SDGs to application-specific demands. When executing a provisioning workflow, the SDG manager decides compatible SDG images on a set of devices selected by the API manager. The Deployment Handler sends the selected SDG images to the Provisioning Daemon in each device that then starts the SGD and configures its virtual environment. Finally, the Provisioning Agent receives a specific application image from the Provisioning Daemon and installs and executes the image.

\section{Algorithms}
\label{sec:algorithms}
\begin{figure}
	\centering
	\includegraphics[width=0.5\textwidth]{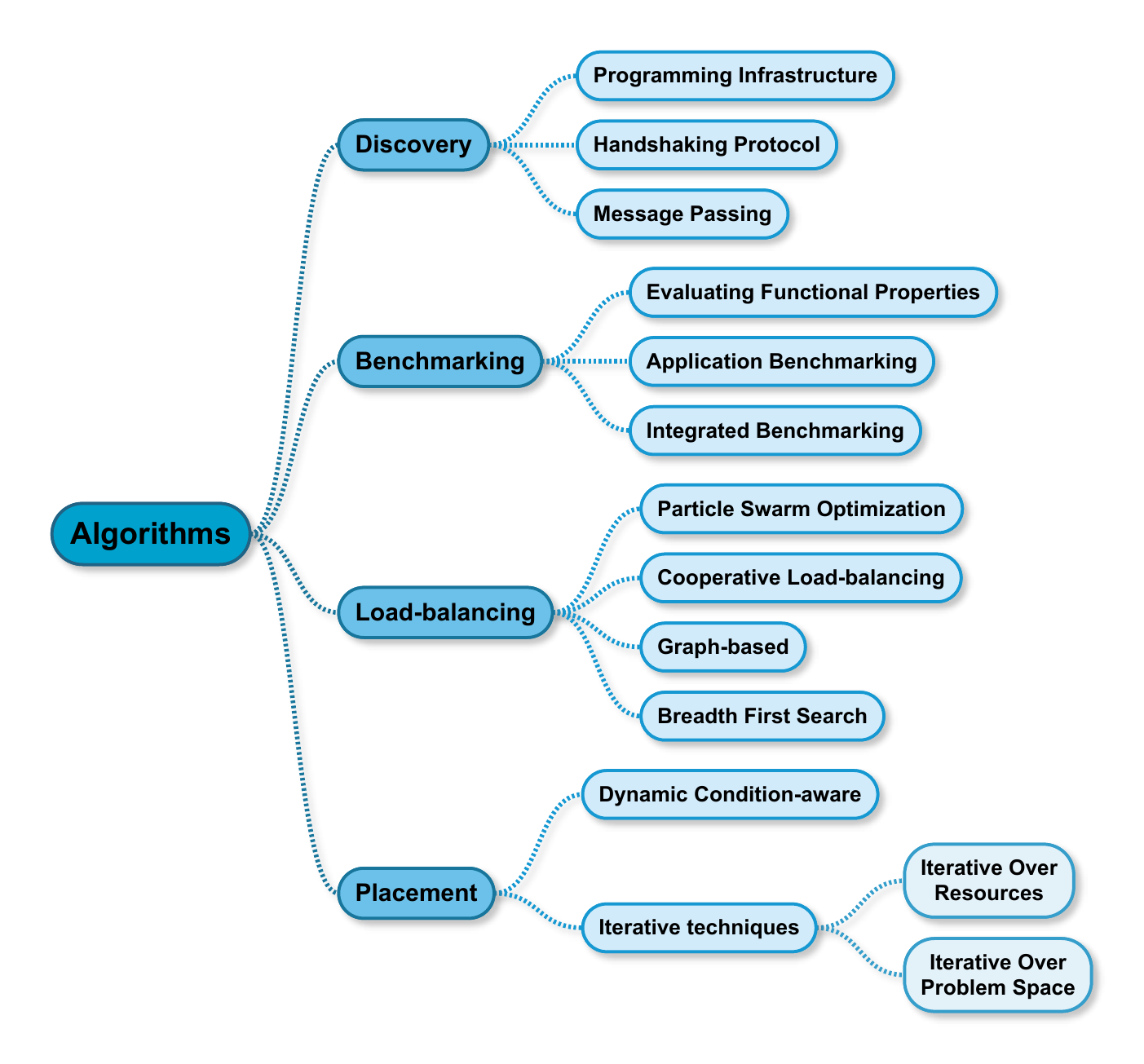}
	\caption{A classification of algorithms in fog/edge computing}
	\label{fig:algo}
\end{figure}

\begin{figure}
	\centering
	\includegraphics[width=0.48\textwidth]{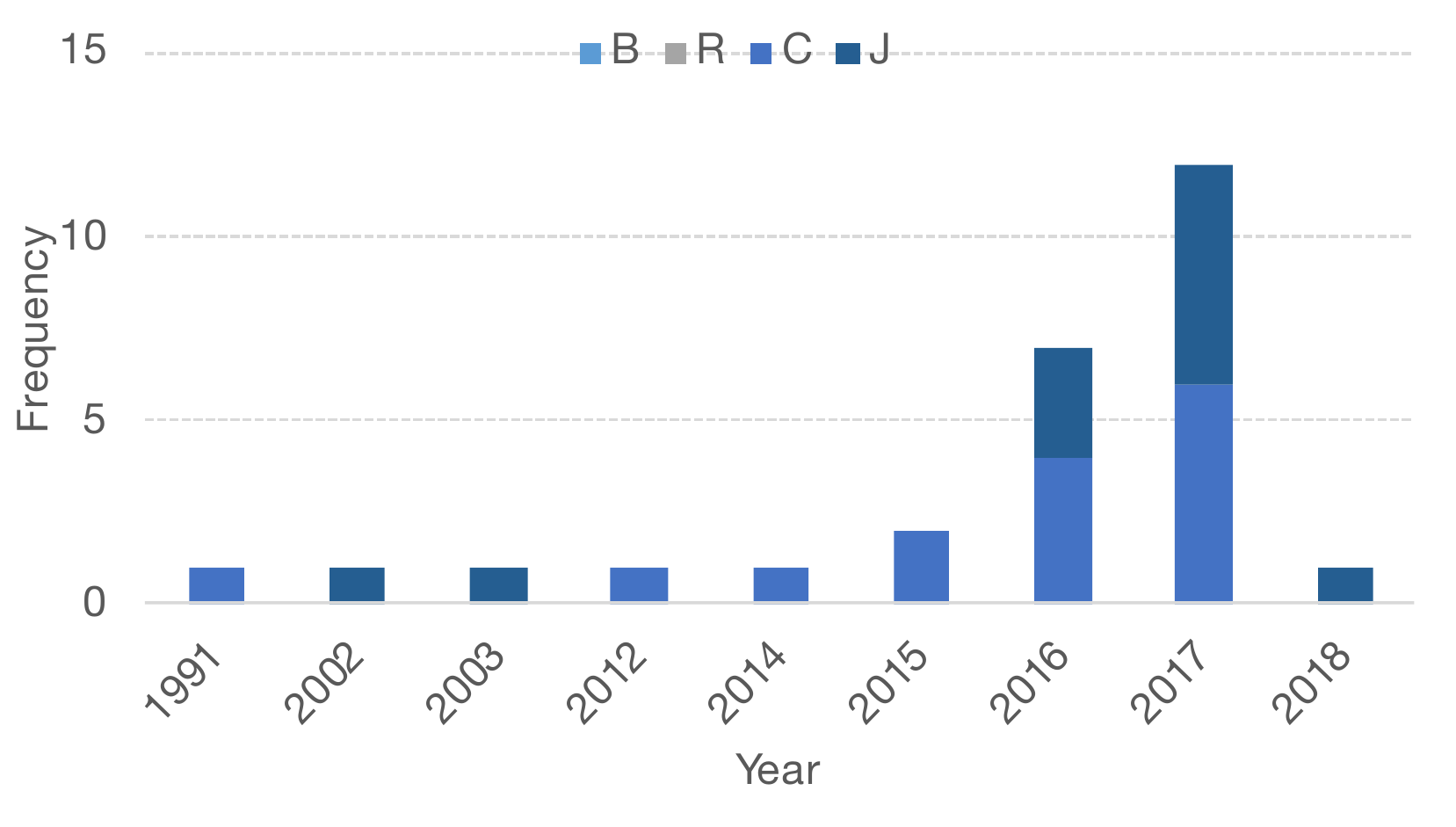}
	\caption{A histogram of publications reviewed for the classification of the algorithms employed for resource management in fog/edge computing. Legend: B - books or book chapters; R - reports, including articles available on pre-print servers or white papers; C - conference or workshop papers; J - journal or magazine articles.}
	\label{fig:histogram-algorithms}
\end{figure}

There are several underlying algorithms used to facilitate fog/edge computing. In this section, we discuss four algorithms, namely (i) discovery~-~identifying edge resources within the network that can be used for distributed computation, (ii) benchmarking~-~capturing the performance of resources for decision-making to maximize the performance of deployments, (iii) load-balancing~-~distributing workloads across resources based on different criteria such as priorities, fairness, etc, and (iv) placement~-~identifying resources appropriate for deploying a workload. Figure \ref{fig:algo} denotes this classification. A histogram of the research publications used is shown in Figure~\ref{fig:histogram-algorithms}.

\subsection{Discovery}
Discovery refers to identifying edge resources so that workloads from the clouds or from user devices/sensors can be deployed on them. Typically, edge computing research assumes that edge resources are discovered. However, this is not an easy task~\cite{disc-1}. Three techniques that use programming infrastructure, handshaking protocols, and message passing are employed in discovery.

The first technique uses \textit{programming infrastructure} such as Foglets, proposed as a mechanism for edge resources to join a cloud-edge ecosystem~\cite{disc-3}. A discovery protocol was proposed that matches the resource requirements of an application against available resources on the edge. Nonetheless, the protocol assumes that the edge resource is publicly known or available for use. An additional join protocol is implemented that allows the selection of one edge node from among a set of resources that have the same geographic distance from the user.

The second technique uses \textit{handshaking protocols}.
The Edge-as-a-Service (EaaS) platform presents a lightweight discovery protocol for a collection of homogeneous edge resources~\cite{disc-2}. The platform requires a master node that may be a compute available network device or a dedicated node that executes a manager process and communicates with edge nodes. The manager communicates with potential edge nodes and executes a process on the edge node to run commands. Once discovered, the Docker or LXD containers can be deployed on edge nodes.

The benefit of the EaaS platform is that the discovery protocol implemented is lightweight and the overhead is only a few seconds for launching, starting, stopping, or terminating containers. Up to 50 containers with an online game workload similar to PokeMon Go were launched on an individual edge node. However, this has been carried out in the context of a single collection of edge nodes. Further research will be required to employ such a model in a federated edge environment. The major drawback of the EaaS platform is that it assumes a centralized master node that can communicate with all potential edge nodes. The handshaking protocol assumes that the edge nodes can be queried and can be made available in a common marketplace via owners. In addition, the security-related implications of the master node installing a manager on the edge node and executing commands on it was not considered.

The third technique for discovery uses \textit{message passing}. In the context of a sensor network in which the end devices may not necessarily have access to the Internet, there is research suggesting that messages may be delivered in such a network using services offered by the nodes (referred to as processing nodes) connected to the Internet~\cite{disc-4}. A discovery method for identifying the processing nodes was presented. The research assumed that a user can communicate with any node in a network and submit queries, and relies on simulation-based validation.

\subsection{Benchmarking}
Benchmarking is a de facto approach for capturing the performance (of entities such as memory, CPU, storage, network, etc) of a computing system~\cite{bench-1}. Metrics relevant to the performance of each entity need to be captured using standard performance evaluation tools. Typical tools used for clusters or supercomputers include LINPACK~\cite{bench-2} or NAS Parallel Benchmarks~\cite{bench-3}.

On the cloud, this is performed by running sample micro or macro applications that stress-tests each entity to obtain a snapshot of the performance of a Virtual Machine (VM) at any given point in time~\cite{bench-4, bench-5}. The key challenge of benchmarking a dynamic computing system (where workloads and conditions change significantly, such as the cloud and the edge) is obtaining metrics in near real-time~\cite{bench-1, bench-6}. Existing benchmarking techniques for the cloud are time-consuming and are not practical solutions because they incur a lot of monetary costs. For example, accurately benchmarking a VM with 200 GB RAM and 1 TB storage requires a few hours. Alternate lightweight benchmarking techniques using containers have been proposed that can obtain results more quickly on the cloud than traditional techniques ~\cite{bench-7, bench-8}. However, a few minutes are still required to get results comparable to traditional benchmarking.

Edge benchmarking can be classified into: (i) benchmarking for evaluating functional properties, (ii) application-based benchmarking, and (iii) integrated benchmarking. The majority of edge benchmarking research evaluates power, CPU, and memory performance of edge processors~\cite{bench-9}.

Benchmarking becomes more challenging in an edge environment for a number of reasons. First, because edge-specific \textit{application benchmarks} that capture a variety of workloads are not yet available. Existing benchmarks are typically scientific applications that are less suited for the edge~\cite{bench-10}. Instead, voice-driven benchmarks~\cite{bench-11} and Internet-of-Things (IoT) applications have been used~\cite{bench-12}. Benchmarking object stores in edge environments have also been proposed~\cite{bench-13}.

Second, running additional time-consuming applications on resource constrained edge nodes can be challenging. There is a need for lightweight benchmarking tools for the edge.

Finally, it is not sufficient to merely benchmark edge resources, but an \textit{integrated approach for benchmarking cloud and edge resources} is required~\cite{bench-14}. This will ensure that the performance of all possible combinations of deployments of the application across the cloud and the edge is considered for maximizing overall application performance.

\subsection{Load-Balancing}

As edge data centers are deployed across the network edge, the issue of distributing tasks using an efficient load-balancing algorithm has gained significant attention. Existing load-balancing algorithms at the edge employ four techniques, namely particle swarm optimization, cooperative load balancing, graph-based balancing, and using breadth-first search.

He et al. \cite{he2016novel} proposed the Software Defined Cloud/Fog Networking (SDCFN) architecture for the Internet of Vehicles (IoV). SDCFN allows centralized control for networking between vehicles and helps the middleware to obtain the required information for load balancing. The study adopted \textit{Particle Swarm Optimization~-~Constrained Optimization} (PSO-CO)~\cite{parsopoulos2002particle} for load-balancing to decrease latency and effectively achieve the required quality of service (QoS) for vehicles.

CooLoad \cite{beraldi2017cooperative} proposed a \textit{cooperative load-balancing }model between fog/edge data centers to decrease service suspension time. CooLoad assigns each data center a buffer to receive requests from clients. When the number of items in the buffer is above a certain threshold, incoming requests to the data center are load-balanced to an adjacent data center. This work assumed that the data centers were connected by a high-speed transport for effective load balancing.

Song et al. \cite{ningning2016Fog} pointed out that existing load-balancing algorithms for cloud platforms that operate in a single cluster cannot be directly applied to a dynamic and peer-to-peer fog computing architecture. To realize efficient load-balancing, they abstracted the fog architecture as a \textit{graph model} where each vertex indicates a node, and the graph edge denotes data dependency between tasks. A dynamic graph-repartitioning algorithm that uses previous load-balancing result as input and minimizes the difference between the load-balancing result and the original status was proposed.

Puthal et al. focused on developing an efficient dynamic load-balancing algorithm with an authentication method for edge data centers~\cite{puthal2018secure}. Tasks were assigned to an under-utilized edge data center by applying the \textit{Breadth First Search} (BFS) method. Each data center was modeled using the current load and the maximum capacity used to compute the current load. The authentication method allows the load-balancing algorithm to find an authenticated data center.

\subsection{Placement}

One challenging issue in fog/edge computing is to place incoming computation tasks on suitable fog/edge resources. Placement algorithms address this issue and need to consider the availability of resources in the fog/edge layer and the environmental changes~\cite{dastjerdi2016Fog}. Existing techniques can be classified as dynamic condition-aware techniques and iterative techniques. Iterative techniques can be further divided into two spaces: iterative over resources, and iterative over the problem spaces.

Wang et al. pointed out that existing work solved placement issues in fog/edge computing under static network conditions and predetermined resource demands and were not \textit{dynamic condition-aware} (do not consider users' mobility and changes in resource availability)~\cite{wang2017dynamic}. This shortcoming was addressed by considering an additional set of parameters including the location and preference of a user, database location, and the load on the system. A method that predicts the values of the parameters when service instances of a user are placed in a certain configuration was proposed. The predicted values yielded an expected cost and optimal placement configuration with lowest cost.

\textit{Iterative methods over resources in the fog computing hierarchy} is another effective technique.
Taneja et al. \cite{taneja2017resource} proposed a placement algorithm for hierarchical fog computing that exploits both conventional cloud and recent fog computing. The algorithm iterates from the fog towards the cloud for placing computation modules first on the available fog nodes. In this algorithm, a node is represented as a set of three attributes: the CPU, memory, and network bandwidth. Each computation module expresses its requirement in the form of the three attributes. The proposed solution first sorts the nodes and modules in ascending order to respectively associate the provided capacity with the requirement. The algorithm then places each module on an appropriate node that has enough resources, iterating from fog nodes to cloud nodes. The authors validated this algorithm using iFogSim, a fog computing simulation toolkit developed by Gupta et al. \cite{gupta2017iFogsim}.

In contrast to the above iterative method, \textit{multiple iterations can be performed over the identified problem space}.
Skarlat et al. proposed an approach called the Fog Service Placement Problem (FSPP) to optimally share resources in fog nodes among IoT services~\cite{skarlat2017towards}. The FSPP considers QoS constraints such as latency or deadline requirements during placement. In the FSPP, a fog node is characterized by three attributes, the CPU, memory, and storage, similar to the work of Taneja et al. \cite{taneja2017resource}. The FSPP suggests a proactive approach where the placement is performed periodically to meet the QoS requirement. When the response time of an application reaches the upper bound, the FSPP prioritizes the application and places it on a node that has enough resources. If there are not enough resources, the algorithm sends the service to the nearest fog network or cloud. The proposed model was evaluated on an extended iFogSim \cite{gupta2017iFogsim}.

\section{Conclusions}
\label{sec:conclusions}
In this survey, we noted that technical challenges to managing the limited resources in fog/edge computing have been addressed to a high degree. However, a few challenges still remain to be made to improve resource management in terms of the capabilities and performance of fog/edge computing. We discuss some future research directions to address the remaining challenges.

Fog/edge computing often employs resource-limited devices such as WiFi APs and set-top boxes that are not suitable for running heavyweight data processing tools such as Apache Spark and deep learning libraries. An alternative lightweight data processing tool such as Apache Quarks can be employed in resource-limited edge devices, but it lacks advanced data analytics functions. The imbalance between lightweight implementations and high performance needs to be addressed.

In fog/edge computing, containers are widely used because they realize lightweight virtualization. However, efficient GPU resource management in containers has not been explored sufficiently, compared to research in virtual machines~\cite{hong2017gpu}. In fog/edge devices, GPUs can be used for data analytics and to assist deep learning algorithms. For example, NVIDIA Jetson TX2 that is a single board computer that hosts an NVIDIA Pascal GPU can be used in the edge with containers. Efficiently managing GPU resources in containers is an open research challenge.

Fog/edge computing has gained significant attention over the last few years as an alternative approach to the conventional centralized cloud computing model. It brings computing resources close to mobile and IoT devices to reduce communication latency and enable efficient use of the network bandwidth. In this survey paper, research on resource management techniques in fog/edge computing was studied to identify and classify the key contributions in the three areas of architectures, infrastructure, and algorithms.


\bibliographystyle{IEEEtran}
\bibliography{references}

\begin{thebibliography}{100}
\providecommand{\url}[1]{#1}
\csname url@samestyle\endcsname
\providecommand{\newblock}{\relax}
\providecommand{\bibinfo}[2]{#2}
\providecommand{\BIBentrySTDinterwordspacing}{\spaceskip=0pt\relax}
\providecommand{\BIBentryALTinterwordstretchfactor}{4}
\providecommand{\BIBentryALTinterwordspacing}{\spaceskip=\fontdimen2\font plus
\BIBentryALTinterwordstretchfactor\fontdimen3\font minus
  \fontdimen4\font\relax}
\providecommand{\BIBforeignlanguage}[2]{{%
\expandafter\ifx\csname l@#1\endcsname\relax
\typeout{** WARNING: IEEEtran.bst: No hyphenation pattern has been}%
\typeout{** loaded for the language `#1'. Using the pattern for}%
\typeout{** the default language instead.}%
\else
\language=\csname l@#1\endcsname
\fi
#2}}
\providecommand{\BIBdecl}{\relax}
\BIBdecl

\bibitem{nextgen-1}
B.~Varghese and R.~Buyya, ``{Next Generation Cloud Computing: New Trends and
  Research Directions},'' \emph{Future Generation Computer Systems}, vol.~79,
  pp. 849 -- 861, 2018.

\bibitem{disc-1}
B.~Varghese, N.~Wang, S.~Barbhuiya, P.~Kilpatrick, and D.~S. Nikolopoulos,
  ``{Challenges and Opportunities in Edge Computing},'' in \emph{IEEE
  International Conference on Smart Cloud}, 2016, pp. 20--26.

\bibitem{intro-2}
W.~Shi, J.~Cao, Q.~Zhang, Y.~Li, and L.~Xu, ``{Edge Computing: Vision and
  Challenges},'' \emph{IEEE Internet of Things Journal}, vol.~3, no.~5, pp.
  637--646, 2016.

\bibitem{satyanarayanan2009case}
M.~Satyanarayanan, P.~Bahl, R.~Caceres, and N.~Davies, ``{The Case for VM-based
  Cloudlets in Mobile Computing},'' \emph{IEEE pervasive Computing}, vol.~8,
  no.~4, 2009.

\bibitem{intro-3}
M.~Satyanarayanan, ``{The Emergence of Edge Computing},'' \emph{Computer},
  vol.~50, no.~1, pp. 30--39, 2017.

\bibitem{intro-4}
F.~Bonomi, R.~Milito, J.~Zhu, and S.~Addepalli, ``{Fog Computing and Its Role
  in the Internet of Things},'' in \emph{Proceedings of the First Edition of
  the MCC Workshop on Mobile Cloud Computing}, 2012, pp. 13--16.

\bibitem{intro-5}
A.~V. Dastjerdi and R.~Buyya, ``{Fog Computing: Helping the Internet of Things
  Realize Its Potential},'' \emph{Computer}, vol.~49, no.~8, pp. 112--116,
  2016.

\bibitem{agg-1}
R.~Rajagopalan and P.~K. Varshney, ``{Data-aggregation Techniques in Sensor
  Networks: A Survey},'' \emph{IEEE Communications Surveys Tutorials}, vol.~8,
  no.~4, pp. 48--63, 2006.

\bibitem{agg-2}
M.~D. de~Assunção, A.~da~Silva~Veith, and R.~Buyya, ``{Distributed Data
  Stream Processing and Edge Computing: A Survey on Resource Elasticity and
  Future Directions},'' \emph{Journal of Network and Computer Applications},
  vol. 103, pp. 1 -- 17, 2018.

\bibitem{agg-3}
E.~Fasolo, M.~Rossi, J.~Widmer, and M.~Zorzi, ``{In-network Aggregation
  Techniques for Wireless Sensor Networks: A Survey},'' \emph{IEEE Wireless
  Communications}, vol.~14, no.~2, pp. 70--87, April 2007.

\bibitem{agg-4}
J.~He, S.~Ji, Y.~Pan, and Y.~Li, ``{Constructing Load-Balanced Data Aggregation
  Trees in Probabilistic Wireless Sensor Networks},'' \emph{IEEE Transactions
  on Parallel and Distributed Systems}, vol.~25, no.~7, pp. 1681--1690, July
  2014.

\bibitem{agg-5}
J.~Tang, Z.~Zhou, J.~Niu, and Q.~Wang, ``{EGF-Tree: An Energy Efficient Index
  Tree for Facilitating Multi-region Query Aggregation in the Internet of
  Things},'' in \emph{IEEE International Conference on Green Computing and
  Communications and IEEE Internet of Things and IEEE Cyber, Physical and
  Social Computing}, 2013, pp. 370--377.

\bibitem{agg-6}
N.~K. Giang, M.~Blackstock, R.~Lea, and V.~C.~M. Leung, ``{Developing IoT
  applications in the Fog: A Distributed Dataflow approach},'' in \emph{{5th
  International Conference on the Internet of Things}}, 2015, pp. 155--162.

\bibitem{agg-12}
Y.~Xu, V.~Mahendran, and S.~Radhakrishnan, ``{Towards SDN-based Fog Computing:
  MQTT Broker Virtualization for Effective and Reliable Delivery},'' in
  \emph{8th International Conference on Communication Systems and Networks},
  2016, pp. 1--6.

\bibitem{agg-7}
H.~Jiang, S.~Jin, and C.~Wang, ``{Prediction or Not? An Energy-Efficient
  Framework for Clustering-Based Data Collection in Wireless Sensor
  Networks},'' \emph{IEEE Transactions on Parallel and Distributed Systems},
  vol.~22, no.~6, pp. 1064--1071, 2011.

\bibitem{agg-8}
F.~Yuan, Y.~Zhan, and Y.~Wang, ``{Data Density Correlation Degree Clustering
  Method for Data Aggregation in WSN},'' \emph{IEEE Sensors Journal}, vol.~14,
  no.~4, pp. 1089--1098, 2014.

\bibitem{agg-9}
M.~Habib~ur Rehman, P.~P. Jayaraman, S.~u.~R. Malik, A.~u.~R. Khan, and
  M.~Medhat~Gaber, ``{RedEdge: A Novel Architecture for Big Data Processing in
  Mobile Edge Computing Environments},'' \emph{Journal of Sensor and Actuator
  Networks}, vol.~6, no.~3, 2017.

\bibitem{agg-10}
S.~Nath, P.~B. Gibbons, S.~Seshan, and Z.~R. Anderson, ``{Synopsis Diffusion
  for Robust Aggregation in Sensor Networks},'' in \emph{Proceedings of the 2nd
  International Conference on Embedded Networked Sensor Systems}, 2004, pp.
  250--262.

\bibitem{agg-11}
S.~Roy, M.~Conti, S.~Setia, and S.~Jajodia, ``{Secure Data Aggregation in
  Wireless Sensor Networks: Filtering out the Attacker's Impact},'' \emph{IEEE
  Transactions on Information Forensics and Security}, vol.~9, no.~4, pp.
  681--694, 2014.

\bibitem{agg-13}
\BIBentryALTinterwordspacing
N.~Zhang, P.~Yang, S.~Zhang, D.~Chen, W.~Zhuang, B.~Liang, and X.~Shen,
  ``{Software Defined Networking Enabled Wireless Network Virtualization:
  Challenges and Solutions},'' \emph{CoRR}, vol. abs/1704.01247, 2017.
  [Online]. Available: \url{http://arxiv.org/abs/1704.01247}
\BIBentrySTDinterwordspacing

\bibitem{agg-14}
M.~Zaharia, T.~Das, H.~Li, T.~Hunter, S.~Shenker, and I.~Stoica, ``{Discretized
  Streams: Fault-tolerant Streaming Computation at Scale},'' in
  \emph{Proceedings of the 24th ACM Symposium on Operating Systems Principles},
  2013, pp. 423--438.

\bibitem{agg-15}
A.~Manjhi, S.~Nath, and P.~B. Gibbons, ``{Tributaries and Deltas: Efficient and
  Robust Aggregation in Sensor Network Streams},'' in \emph{Proceedings of the
  ACM SIGMOD International Conference on Management of Data}, 2005, pp.
  287--298.

\bibitem{agg-16}
D.~S. Mantri, N.~R. Prasad, and R.~Prasad, ``{Bandwidth Efficient Cluster-based
  Data Aggregation for Wireless Sensor Network},'' \emph{Computers and
  Electrical Engineering}, vol.~41, no.~C, pp. 256--264, Jan. 2015.

\bibitem{agg-17}
L.~Becchetti, P.~Korteweg, A.~Marchetti-Spaccamela, M.~Skutella, L.~Stougie,
  and A.~Vitaletti, ``{Latency Constrained Aggregation in Sensor Networks},''
  in \emph{Algorithms -- ESA 2006}, Y.~Azar and T.~Erlebach, Eds.\hskip 1em
  plus 0.5em minus 0.4em\relax Berlin, Heidelberg: Springer Berlin Heidelberg,
  2006, pp. 88--99.

\bibitem{agg-18}
H.~Li, C.~Wu, Q.-S. Hua, and F.~C.~M. Lau, ``{Latency-minimizing Data
  Aggregation in Wireless Sensor Networks Under Physical Interference Model},''
  \emph{Ad Hoc Networks}, vol.~12, pp. 52--68, Jan. 2014.

\bibitem{agg-19}
S.~Reiff-Marganiec, M.~Tilly, and H.~Janicke, ``{Low-Latency Service Data
  Aggregation Using Policy Obligations},'' in \emph{IEEE International
  Conference on Web Services}, 2014, pp. 526--533.

\bibitem{agg-20}
H.~Li, C.~Wu, D.~Yu, Q.~S. Hua, and F.~C.~M. Lau, ``{Aggregation Latency-Energy
  Trade-off in Wireless Sensor Networks with Successive Interference
  Cancellation},'' \emph{IEEE Transactions on Parallel and Distributed
  Systems}, vol.~24, no.~11, pp. 2160--2170, 2013.

\bibitem{agg-21}
C.~Anagnostopoulos, ``Time-optimized contextual information forwarding in
  mobile sensor networks,'' \emph{Journal of Parallel and Distributed
  Computing}, vol.~74, no.~5, pp. 2317--2332, 2014.

\bibitem{agg-22}
N.~Harth and C.~Anagnostopoulos, ``{Quality-aware Aggregation and Predictive
  Analytics at the Edge},'' in \emph{IEEE International Conference on Big
  Data}, 2017, pp. 17--26.

\bibitem{agg-23}
H.~Wang, Z.~Wang, and J.~Domingo-Ferrer, ``{Anonymous and Secure Aggregation
  Scheme in Fog-based Public Cloud Computing},'' \emph{Future Generation
  Computer Systems}, vol.~78, pp. 712 -- 719, 2018.

\bibitem{agg-24}
G.~Castagnos and F.~Laguillaumie, ``{Linearly Homomorphic Encryption from
  DDH},'' in \emph{Topics in Cryptology --- CT-RSA 2015}, K.~Nyberg, Ed.\hskip
  1em plus 0.5em minus 0.4em\relax Springer International Publishing, 2015, pp.
  487--505.

\bibitem{agg-25}
R.~Lu, K.~Heung, A.~H. Lashkari, and A.~A. Ghorbani, ``{A Lightweight
  Privacy-Preserving Data Aggregation Scheme for Fog Computing-Enhanced IoT},''
  \emph{IEEE Access}, vol.~5, pp. 3302--3312, 2017.

\bibitem{agg-25a}
Y.~Chen, Z.~Lu, H.~Xiong, and W.~Xu, ``{Privacy-Preserving Data Aggregation
  Protocol for Fog Computing-Assisted Vehicle-to-Infrastructure Scenario},''
  \emph{Security and Communication Networks}, p.~14, 2018.

\bibitem{agg-26}
\BIBentryALTinterwordspacing
B.~Varghese, N.~Wang, D.~S. Nikolopoulos, and R.~Buyya, ``{Feasibility of Fog
  Computing},'' \emph{CoRR}, vol. abs/1701.05451, 2017. [Online]. Available:
  \url{http://arxiv.org/abs/1701.05451}
\BIBentrySTDinterwordspacing

\bibitem{agg-27}
D.~Mantri, N.~R. Prasad, and R.~Prasad, ``{BHCDA: Bandwidth efficient
  heterogeneity aware cluster based data aggregation for Wireless Sensor
  Network},'' in \emph{International Conference on Advances in Computing,
  Communications and Informatics}, 2013, pp. 1064--1069.

\bibitem{agg-28}
D.~S. Mantri, N.~R. Prasad, and R.~Prasad, ``{Mobility and Heterogeneity Aware
  Cluster-Based Data Aggregation for Wireless Sensor Network},'' \emph{Wireless
  Personal Communications}, vol.~86, no.~2, pp. 975--993, Jan 2016.

\bibitem{agg-29}
Y.~Xiao, M.~Noreikis, and A.~Ylä-Jaäiski, ``{QoS-oriented Capacity Planning
  for Edge Computing},'' in \emph{IEEE International Conference on
  Communications}, 2017, pp. 1--6.

\bibitem{sharing-1}
H.~T. Dinh, C.~Lee, D.~Niyato, and P.~Wang, ``{A Survey of Mobile Cloud
  Computing: Architecture, Applications, and Approaches},'' \emph{Wireless
  Communications and Mobile Computing}, vol.~13, no.~18, pp. 1587--1611, 2013.

\bibitem{sharing-2}
Y.~Mao, C.~You, J.~Zhang, K.~Huang, and K.~B. Letaief, ``{A Survey on Mobile
  Edge Computing: The Communication Perspective},'' \emph{IEEE Communications
  Surveys Tutorials}, vol.~19, no.~4, pp. 2322--2358, 2017.

\bibitem{sharing-3}
D.~Milojicic, V.~Kalogeraki, R.~Lukose, K.~Nagaraja, J.~Pruyne, and B.~Richard,
  ``{Peer-to-Peer Computing},'' HP Laboratories, Tech. Rep. HPL-2002-57, March
  2002.

\bibitem{sharing-4}
K.~Habak, M.~Ammar, K.~A. Harras, and E.~Zegura, ``{Femto Clouds: Leveraging
  Mobile Devices to Provide Cloud Service at the Edge},'' in \emph{Proceedings
  of the IEEE 8th International Conference on Cloud Computing}, 2015, pp.
  9--16.

\bibitem{sharing-5}
K.~Habak, E.~W. Zegura, M.~H. Ammar, and K.~A. Harras, ``{Workload Management
  for Dynamic Mobile Device Clusters in Edge Femtoclouds},'' in
  \emph{Proceedings of the 2nd {ACM/IEEE} Symposium on Edge Computing}, 2017,
  pp. 6:1--6:14.

\bibitem{sharing-6}
Y.~Cui, J.~Song, K.~Ren, M.~Li, Z.~Li, Q.~Ren, and Y.~Zhang, ``{Software
  Defined Cooperative Offloading for Mobile Cloudlets},'' \emph{IEEE/ACM
  Transactions on Networking}, vol.~25, no.~3, pp. 1746--1760, 2017.

\bibitem{sharing-7}
X.~Chen, ``{Decentralized Computation Offloading Game for Mobile Cloud
  Computing},'' \emph{IEEE Transactions on Parallel and Distributed Systems},
  vol.~26, no.~4, pp. 974--983, 2015.

\bibitem{sharing-8}
X.~Ma, C.~Lin, X.~Xiang, and C.~Chen, ``{Game-theoretic Analysis of Computation
  Offloading for Cloudlet-based Mobile Cloud Computing},'' in \emph{Proceedings
  of the 18th ACM International Conference on Modelling, Analysis and
  Simulation of Wireless and Mobile Systems}, 2015, pp. 271--278.

\bibitem{sharing-9}
W.~Gao, ``{Opportunistic Peer-to-Peer Mobile Cloud Computing at the Tactical
  Edge},'' in \emph{IEEE Military Communications Conference}, Oct 2014, pp.
  1614--1620.

\bibitem{sharing-10}
A.~Mtibaa, A.~Fahim, K.~A. Harras, and M.~H. Ammar, ``{Towards Resource Sharing
  in Mobile Device Clouds: Power Balancing Across Mobile Devices},'' \emph{ACM
  SIGCOMM Computer Communication Review}, vol.~43, no.~4, pp. 51--56, Aug.
  2013.

\bibitem{sharing-11}
C.~Funai, C.~Tapparello, and W.~Heinzelman, ``{Mobile to Mobile Computational
  Offloading in Multi-Hop Cooperative Networks},'' in \emph{IEEE Global
  Communications Conference}, 2016, pp. 1--7.

\bibitem{sharing-12}
C.~Shi, V.~Lakafosis, M.~H. Ammar, and E.~W. Zegura, ``{Serendipity: Enabling
  Remote Computing Among Intermittently Connected Mobile Devices},'' in
  \emph{Proceedings of the 13th ACM International Symposium on Mobile Ad Hoc
  Networking and Computing}, 2012, pp. 145--154.

\bibitem{sharing-13}
E.~Borgia, R.~Bruno, M.~Conti, D.~Mascitti, and A.~Passarella, ``{Mobile Edge
  Clouds for Information-Centric IoT Services},'' in \emph{IEEE Symposium on
  Computers and Communication}, 2016, pp. 422--428.

\bibitem{sharing-14}
Z.~Sanaei, S.~Abolfazli, A.~Gani, and R.~Buyya, ``{Heterogeneity in Mobile
  Cloud Computing: Taxonomy and Open Challenges},'' \emph{IEEE Communications
  Surveys Tutorials}, vol.~16, no.~1, pp. 369--392, 2014.

\bibitem{sharing-15}
N.~Fernando, S.~W. Loke, and W.~Rahayu, ``{Computing with Nearby Mobile
  Devices: a Work Sharing Algorithm for Mobile Edge-Clouds},'' \emph{IEEE
  Transactions on Cloud Computing}, vol.~PP, no.~99, pp. 1--1, 2017.

\bibitem{sharing-16}
T.~Nishio, R.~Shinkuma, T.~Takahashi, and N.~B. Mandayam, ``{Service-oriented
  Heterogeneous Resource Sharing for Optimizing Service Latency in Mobile
  Cloud},'' in \emph{Proceedings of the 1st International Workshop on Mobile
  Cloud Computing and Networking}, 2013, pp. 19--26.

\bibitem{sharing-17}
A.~Mtibaa, K.~Harras, and H.~Alnuweiri, ``{Friend or Foe? Detecting and
  Isolating Malicious Nodes in Mobile Edge Computing Platforms},'' in \emph{7th
  International Conference on Cloud Computing Technology and Science
  (CloudCom)}, 2015, pp. 42--49.

\bibitem{sharing-18}
H.~Viswanathan, E.~K. Lee, and D.~Pompili, ``{A Multi-Objective Approach to
  Real-Time In-Situ Processing of Mobile-Application Workflows},'' \emph{IEEE
  Transactions on Parallel and Distributed Systems}, vol.~27, no.~11, pp.
  3116--3130, 2016.

\bibitem{sharing-19}
J.~Panneerselvam, J.~Hardy, L.~Liu, B.~Yuan, and N.~Antonopoulos, ``{Mobilouds:
  An Energy Efficient MCC Collaborative Framework With Extended Mobile
  Participation for Next Generation Networks},'' \emph{IEEE Access}, vol.~4,
  pp. 9129--9144, 2016.

\bibitem{sharing-20}
T.~Penner, A.~Johnson, B.~V. Slyke, M.~Guirguis, and Q.~Gu, ``{Transient
  clouds: Assignment and Collaborative Execution of Tasks on Mobile Devices},''
  in \emph{IEEE Global Communications Conference}, 2014, pp. 2801--2806.

\bibitem{sharing-21}
I.~Farris, L.~Militano, M.~Nitti, L.~Atzori, and A.~Iera, ``{MIFaaS: A
  Mobile-IoT-Federation-as-a-Service Model for Dynamic Cooperation of IoT Cloud
  Providers},'' \emph{Future Generation Computer Systems}, vol.~70, pp.
  126--137, 2017.

\bibitem{off-1}
P.~Simoens, Y.~Xiao, P.~Pillai, Z.~Chen, K.~Ha, and M.~Satyanarayanan,
  ``{Scalable Crowd-sourcing of Video from Mobile Devices},'' in
  \emph{Proceeding of the 11th Annual International Conference on Mobile
  Systems, Applications, and Services}, 2013, pp. 139--152.

\bibitem{off-2}
\BIBentryALTinterwordspacing
L.~Chen, J.~Wu, X.~Long, and Z.~Zhang, ``{{ENGINE:} Cost Effective Offloading
  in Mobile Edge Computing with Fog-Cloud Cooperation},'' \emph{CoRR}, vol.
  abs/1711.01683, 2017. [Online]. Available:
  \url{http://arxiv.org/abs/1711.01683}
\BIBentrySTDinterwordspacing

\bibitem{off-3}
A.~Brogi and S.~Forti, ``{QoS-Aware Deployment of IoT Applications Through the
  Fog},'' \emph{IEEE Internet of Things Journal}, vol.~4, no.~5, pp.
  1185--1192, 2017.

\bibitem{off-3a}
R.~Mahmud, S.~N. Srirama, K.~Ramamohanarao, and R.~Buyya, ``Quality of
  experience (qoe)-aware placement of applications in fog computing
  environments,'' \emph{Journal of Parallel and Distributed Computing}, 2018.

\bibitem{off-4}
B.-G. Chun, S.~Ihm, P.~Maniatis, M.~Naik, and A.~Patti, ``{CloneCloud: Elastic
  Execution Between Mobile Device and Cloud},'' in \emph{Proceedings of the
  Sixth Conference on Computer Systems}, 2011, pp. 301--314.

\bibitem{off-5}
A.~Bhattcharya and P.~De, ``{Computation Offloading from Mobile Devices: Can
  Edge Devices Perform Better Than the Cloud?}'' in \emph{Proceedings of the
  Third International Workshop on Adaptive Resource Management and Scheduling
  for Cloud Computing}, 2016, pp. 1--6.

\bibitem{off-6}
N.~Takahashi, H.~Tanaka, and R.~Kawamura, ``{Analysis of Process Assignment in
  Multi-tier mobile Cloud Computing and Application to Edge Accelerated Web
  Browsing},'' in \emph{3rd IEEE International Conference on Mobile Cloud
  Computing, Services, and Engineering}, 2015, pp. 233--234.

\bibitem{off-7}
Y.~Kang, J.~Hauswald, C.~Gao, A.~Rovinski, T.~Mudge, J.~Mars, and L.~Tang,
  ``{Neurosurgeon: Collaborative Intelligence Between the Cloud and Mobile
  Edge},'' in \emph{Proceedings of the Twenty-Second International Conference
  on Architectural Support for Programming Languages and Operating Systems},
  2017, pp. 615--629.

\bibitem{off-8}
S.~Teerapittayanon, B.~McDanel, and H.~T. Kung, ``{Distributed Deep Neural
  Networks Over the Cloud, the Edge and End Devices},'' in \emph{IEEE 37th
  International Conference on Distributed Computing Systems}, 2017, pp.
  328--339.

\bibitem{off-9}
K.~Tokunaga, K.~Kawamura, and N.~Takaya, ``{High-speed Uploading Architecture
  using Distributed Edge Servers on Multi-RAT Heterogeneous Networks},'' in
  \emph{IEEE International Symposium on Local and Metropolitan Area Networks},
  2016, pp. 1--2.

\bibitem{off-10}
Q.~Xu, Z.~Su, Q.~Zheng, M.~Luo, and B.~Dong, ``{Secure Content Delivery with
  Edge Nodes to Save Caching Resources for Mobile Users in Green Cities},''
  \emph{IEEE Transactions on Industrial Informatics}, vol.~PP, no.~99, pp.
  1--1, 2017.

\bibitem{off-11}
Y.~Lin, B.~Kemme, M.~Patino-Martinez, and R.~Jimenez-Peris, ``{Enhancing Edge
  Computing with Database Replication},'' in \emph{26th IEEE International
  Symposium on Reliable Distributed Systems (SRDS 2007)}, 2007, pp. 45--54.

\bibitem{off-12}
L.~Gao, M.~Dahlin, A.~Nayate, J.~Zheng, and A.~Iyengar, ``{Application Specific
  Data Replication for Edge Services},'' in \emph{Proceedings of the 12th
  International Conference on World Wide Web}, 2003, pp. 449--460.

\bibitem{off-13}
Z.~Chen, L.~Jiang, W.~Hu, K.~Ha, B.~Amos, P.~Pillai, A.~Hauptmann, and
  M.~Satyanarayanan, ``{Early Implementation Experience with Wearable Cognitive
  Assistance Applications},'' in \emph{Proceedings of the Workshop on Wearable
  Systems and Applications}, 2015, pp. 33--38.

\bibitem{off-14}
N.~Wang, B.~Varghese, M.~Matthaiou, and D.~S. Nikolopoulos, ``{ENORM: A
  Framework For Edge NOde Resource Management},'' \emph{IEEE Transactions on
  Services Computing}, vol.~PP, no.~99, pp. 1--1, 2017.

\bibitem{off-19}
M.~Báguena, G.~Samaras, A.~Pamboris, M.~L. Sichitiu, P.~Pietzuch, and
  P.~Manzoni, ``{Towards Enabling Hyper-responsive Mobile Apps Through Network
  Edge Assistance},'' in \emph{13th IEEE Annual Consumer Communications
  Networking Conference}, 2016, pp. 399--404.

\bibitem{off-16}
K.~Bhardwaj, P.~Agrawal, A.~Gavrilovska, and K.~Schwan, ``{AppSachet:
  Distributed App Delivery from the Edge Cloud},'' in \emph{Mobile Computing,
  Applications, and Services - 7th International Conference, MobiCASE 2015,
  Berlin, Germany, November 12-13, 2015, Revised Selected Papers}, 2015, pp.
  89--106.

\bibitem{off-17}
E.~Zeydan, E.~Bastug, M.~Bennis, M.~A. Kader, I.~A. Karatepe, A.~S. Er, and
  M.~Debbah, ``{Big Data Caching for Networking: Moving from Cloud to Edge},''
  \emph{IEEE Communications Magazine}, vol.~54, no.~9, pp. 36--42, 2016.

\bibitem{off-18}
\BIBentryALTinterwordspacing
T.~X. Vu, S.~Chatzinotas, and B.~E. Ottersten, ``{Edge-Caching Wireless
  Networks: Energy-Efficient Design and Optimization},'' \emph{CoRR}, vol.
  abs/1705.05590, 2017. [Online]. Available:
  \url{http://arxiv.org/abs/1705.05590}
\BIBentrySTDinterwordspacing

\bibitem{off-15}
P.~Savolainen, S.~Helal, J.~Reitmaa, K.~Kuikkaniemi, G.~Jacucci, M.~Rinne,
  M.~Turpeinen, and S.~Tarkoma, ``{Spaceify: A Client-edge-server Ecosystem for
  Mobile Computing in Smart Spaces},'' in \emph{Proceedings of the 19th Annual
  International Conference on Mobile Computing and Networking}, 2013, pp.
  211--214.

\bibitem{control-1}
N.~Mohan and J.~Kangasharju, ``{Edge-Fog Cloud: A Distributed Cloud for
  Internet of Things Computations},'' in \emph{Cloudification of the Internet
  of Things}, 2016, pp. 1--6.

\bibitem{control-2}
X.~Chen and J.~Zhang, ``{When D2D Meets Cloud: Hybrid Mobile Task Offloadings
  in Fog Computing},'' in \emph{IEEE International Conference on
  Communications}, 2017, pp. 1--6.

\bibitem{control-3}
A.~Stanciu, ``{Blockchain Based Distributed Control System for Edge
  Computing},'' in \emph{21st International Conference on Control Systems and
  Computer Science (CSCS)}, 2017, pp. 667--671.

\bibitem{control-4}
X.~Chen, L.~Jiao, W.~Li, and X.~Fu, ``{Efficient Multi-User Computation
  Offloading for Mobile-Edge Cloud Computing},'' \emph{IEEE/ACM Transactions on
  Networking}, vol.~24, no.~5, pp. 2795--2808, 2016.

\bibitem{control-5}
Y.~Sahni, J.~Cao, S.~Zhang, and L.~Yang, ``{Edge Mesh: A New Paradigm to Enable
  Distributed Intelligence in Internet of Things},'' \emph{IEEE Access},
  vol.~5, pp. 16\,441--16\,458, 2017.

\bibitem{control-6}
F.~Van~den Abeele, J.~Hoebeke, G.~K. Teklemariam, I.~Moerman, and P.~Demeester,
  ``{Sensor Function Virtualization to Support Distributed Intelligence in the
  Internet of Things},'' \emph{Wireless Personal Communications}, vol.~81,
  no.~4, pp. 1415--1436, Apr 2015.

\bibitem{openfog-1}
``{OpenFog Reference Architecture for Fog Computing},''
  \url{https://www.openfogconsortium.org/wp-content/uploads/OpenFog_Reference_Architecture_2_09_17-FINAL.pdf},
  2017, accessed: 08/03/2018.

\bibitem{virtualization-1}
P.~Liu, D.~Willis, and S.~Banerjee, ``{ParaDrop: Enabling Lightweight
  Multi-tenancy at the Network’s Extreme Edge},'' in \emph{IEEE/ACM Symposium
  on Edge Computing}, Oct 2016, pp. 1--13.

\bibitem{virtualization-2}
R.~Morabito, V.~Cozzolino, A.~Y. Ding, N.~Beijar, and J.~Ott, ``{Consolidate
  IoT Edge Computing with Lightweight Virtualization},'' \emph{IEEE Network},
  vol.~32, no.~1, pp. 102--111, Jan 2018.

\bibitem{slicing-1}
O.~Sallent, J.~Perez-Romero, R.~Ferrus, and R.~Agusti, ``{On Radio Access
  Network Slicing from a Radio Resource Management Perspective},'' \emph{IEEE
  Wireless Communications}, vol.~24, no.~5, pp. 166--174, October 2017.

\bibitem{slicing-2}
V.~Jeyakumar, M.~Alizadeh, D.~Mazi\`{e}res, B.~Prabhakar, C.~Kim, and
  A.~Greenberg, ``{EyeQ: Practical Network Performance Isolation at the
  Edge},'' in \emph{Proceedings of the 10th USENIX Conference on Networked
  Systems Design and Implementation}, 2013, pp. 297--312.

\bibitem{confais2017object}
B.~Confais, A.~Lebre, and B.~Parrein, ``{An Object Store Service for a Fog/Edge
  Computing Infrastructure Based on IPFS and a Scale-Out NAS},'' in \emph{IEEE
  1st International Conference on Fog and Edge Computing}, 2017, pp. 41--50.

\bibitem{stojmenovic2014Fog}
I.~Stojmenovic, ``{Fog Computing: A Cloud to the Ground Support for Smart
  Things and Machine-to-Machine Networks},'' in \emph{Australasian
  Telecommunication Networks and Applications Conference}, 2014, pp. 117--122.

\bibitem{Cox2013}
S.~J. Johnston, P.~J. Basford, C.~S. Perkins, H.~Herry, F.~P. Tso, D.~Pezaros,
  R.~D. Mullins, E.~Yoneki, S.~J. Cox, and J.~Singer, ``{Commodity Single Board
  Computer Clusters and Their Applications},'' \emph{Future Generation Computer
  Systems}, June 2018.

\bibitem{amento2016focusstack}
B.~Amento, B.~Balasubramanian, R.~J. Hall, K.~Joshi, G.~Jung, and K.~H. Purdy,
  ``{FocusStack: Orchestrating Edge Clouds Using Location-Based Focus of
  Attention},'' in \emph{Edge Computing (SEC), IEEE/ACM Symposium on}.\hskip
  1em plus 0.5em minus 0.4em\relax IEEE, 2016, pp. 179--191.

\bibitem{bellavista2017feasibility}
P.~Bellavista and A.~Zanni, ``{Feasibility of Fog Computing Deployment based on
  Docker Containerization over Raspberry Pi},'' in \emph{Proceedings of the
  18th International Conference on Distributed Computing and Networking}.\hskip
  1em plus 0.5em minus 0.4em\relax ACM, 2017, p.~16.

\bibitem{hong2017Cloud}
H.-J. Hong, ``{From Cloud Computing to Fog Computing: Unleash the Power of Edge
  and End Devices},'' in \emph{IEEE International Conference on Cloud Computing
  Technology and Science}.\hskip 1em plus 0.5em minus 0.4em\relax IEEE, 2017,
  pp. 331--334.

\bibitem{hong2016animation}
H.-J. Hong, J.-C. Chuang, and C.-H. Hsu, ``{Animation Rendering on Multimedia
  Fog Computing Platforms},'' in \emph{IEEE International Conference on Cloud
  Computing Technology and Science}.\hskip 1em plus 0.5em minus 0.4em\relax
  IEEE, 2016, pp. 336--343.

\bibitem{aazam2014Fog}
M.~Aazam and E.-N. Huh, ``{Fog Computing and Smart Gateway Based Communication
  for Cloud of Things},'' in \emph{Future Internet of Things and Cloud
  (FiCloud), 2014 International Conference on}.\hskip 1em plus 0.5em minus
  0.4em\relax IEEE, 2014, pp. 464--470.

\bibitem{gosain2016enabling}
A.~Gosain, M.~Berman, M.~Brinn, T.~Mitchell, C.~Li, Y.~Wang, H.~Jin, J.~Hua,
  and H.~Zhang, ``{Enabling Campus Edge Computing Using Geni Racks and Mobile
  Resources},'' in \emph{Edge Computing (SEC), IEEE/ACM Symposium on}, 2016,
  pp. 41--50.

\bibitem{chiueh2005survey}
S.~N. T.-c. Chiueh and S.~Brook, ``A survey on virtualization technologies,''
  \emph{Rpe Report}, vol. 142, 2005.

\bibitem{hong2017gpu}
C.-H. Hong, I.~Spence, and D.~S. Nikolopoulos, ``{GPU Virtualization and
  Scheduling Methods: A Comprehensive Survey},'' \emph{ACM Computing Surveys
  (CSUR)}, vol.~50, no.~3, p.~35, 2017.

\bibitem{barham2003xen}
P.~Barham, B.~Dragovic, K.~Fraser, S.~Hand, T.~Harris, A.~Ho, R.~Neugebauer,
  I.~Pratt, and A.~Warfield, ``{Xen and the Art of Virtualization},'' in
  \emph{ACM SIGOPS operating systems review}, vol.~37, no.~5, 2003, pp.
  164--177.

\bibitem{kivity2007kvm}
A.~Kivity, Y.~Kamay, D.~Laor, U.~Lublin, and A.~Liguori, ``{KVM: The Linux
  Virtual Machine Monitor},'' in \emph{Proceedings of the Linux symposium},
  vol.~1, 2007, pp. 225--230.

\bibitem{gu2017cost}
L.~Gu, D.~Zeng, S.~Guo, A.~Barnawi, and Y.~Xiang, ``{Cost Efficient Resource
  Management in Fog Computing Supported Medical Cyber-Physical System},''
  \emph{IEEE Transactions on Emerging Topics in Computing}, vol.~5, no.~1, pp.
  108--119, 2017.

\bibitem{wang2017elastic}
J.~Wang, J.~Pan, and F.~Esposito, ``{Elastic Urban Video Surveillance System
  Using Edge Computing},'' in \emph{Proceedings of the Workshop on Smart
  Internet of Things}, 2017, p.~7.

\bibitem{lee2012challenges}
I.~Lee, O.~Sokolsky, S.~Chen, J.~Hatcliff, E.~Jee, B.~Kim, A.~King,
  M.~Mullen-Fortino, S.~Park, A.~Roederer \emph{et~al.}, ``{Challenges and
  Research Directions in Medical Cyber--Physical Systems},'' \emph{Proceedings
  of the IEEE}, vol. 100, no.~1, pp. 75--90, 2012.

\bibitem{vaughan2006new}
S.~J. Vaughan-Nichols, ``New approach to virtualization is a lightweight,''
  \emph{Computer}, vol.~39, no.~11, 2006.

\bibitem{haydel2015enhancing}
N.~Haydel, S.~Gesing, I.~Taylor, G.~Madey, A.~Dakkak, S.~G. De~Gonzalo, and
  W.-M.~W. Hwu, ``{Enhancing the Usability and Utilization of Accelerated
  Architectures via Docker},'' in \emph{IEEE/ACM 8th International Conference
  on Utility and Cloud Computing}, 2015, pp. 361--367.

\bibitem{pahl2015containers}
C.~Pahl and B.~Lee, ``{Containers and Clusters for Edge Cloud Architectures--A
  Technology Review},'' in \emph{Future Internet of Things and Cloud (FiCloud),
  2015 3rd International Conference on}.\hskip 1em plus 0.5em minus 0.4em\relax
  IEEE, 2015, pp. 379--386.

\bibitem{helsley2009lxc}
M.~Helsley, ``{LXC: Linux Container Tools},'' \emph{IBM developerWorks
  Technical Library}, vol.~11, 2009.

\bibitem{merkel2014docker}
D.~Merkel, ``{Docker: Lightweight Linux Containers for Consistent Development
  and Deployment},'' \emph{Linux Journal}, vol. 2014, no. 239, p.~2, 2014.

\bibitem{brewer2015kubernetes}
E.~A. Brewer, ``{Kubernetes and the Path to Cloud Native},'' in
  \emph{Proceedings of the Sixth ACM Symposium on Cloud Computing}, 2015, pp.
  167--167.

\bibitem{morabito2016enabling}
R.~Morabito and N.~Beijar, ``{Enabling Data Processing at the Network Edge
  Through Lightweight Virtualization Technologies},'' in \emph{IEEE
  International Conference on Sensing, Communication and Networking}, 2016, pp.
  1--6.

\bibitem{medina2014survey}
V.~Medina and J.~M. Garc{\'\i}a, ``{A Survey of Migration Mechanisms of Virtual
  Machines},'' \emph{ACM Computing Surveys}, vol.~46, no.~3, p.~30, 2014.

\bibitem{ahmad2015survey}
R.~W. Ahmad, A.~Gani, S.~H.~A. Hamid, M.~Shiraz, A.~Yousafzai, and F.~Xia, ``{A
  Survey on Virtual Machine Migration and Server Consolidation Frameworks for
  Cloud Data Centers},'' \emph{Journal of Network and Computer Applications},
  vol.~52, pp. 11--25, 2015.

\bibitem{forsman2015algorithms}
M.~Forsman, A.~Glad, L.~Lundberg, and D.~Ilie, ``{Algorithms for Automated Live
  Migration of Virtual Machines},'' \emph{Journal of Systems and Software},
  vol. 101, pp. 110--126, 2015.

\bibitem{cerroni2014live}
W.~Cerroni and F.~Callegati, ``{Live Migration of Virtual Network Functions in
  Cloud-based Edge Networks},'' in \emph{IEEE International Conference on
  Communications}.\hskip 1em plus 0.5em minus 0.4em\relax IEEE, 2014, pp.
  2963--2968.

\bibitem{shekhar2017indices}
S.~Shekhar, A.~D. Chhokra, A.~Bhattacharjee, G.~Aupy, and A.~Gokhale,
  ``{INDICES: Exploiting Edge Resources for Performance-aware Cloud Hosted
  Services},'' in \emph{IEEE 1st International Conference on Fog and Edge
  Computing}.\hskip 1em plus 0.5em minus 0.4em\relax IEEE, 2017, pp. 75--80.

\bibitem{bittencourt2015towards}
L.~F. Bittencourt, M.~M. Lopes, I.~Petri, and O.~F. Rana, ``{Towards Virtual
  Machine Migration in Fog Computing},'' in \emph{10th International Conference
  on P2P, Parallel, Grid, Cloud and Internet Computing}.\hskip 1em plus 0.5em
  minus 0.4em\relax IEEE, 2015, pp. 1--8.

\bibitem{chowdhury2010survey}
N.~M.~K. Chowdhury and R.~Boutaba, ``A survey of network virtualization,''
  \emph{Computer Networks}, vol.~54, no.~5, pp. 862--876, 2010.

\bibitem{kreutz2015software}
D.~Kreutz, F.~M. Ramos, P.~E. Verissimo, C.~E. Rothenberg, S.~Azodolmolky, and
  S.~Uhlig, ``{Software-Defined Networking: A Comprehensive Survey},''
  \emph{Proceedings of the IEEE}, vol. 103, no.~1, pp. 14--76, 2015.

\bibitem{han2015network}
B.~Han, V.~Gopalakrishnan, L.~Ji, and S.~Lee, ``{Network Function
  Virtualization: Challenges and Opportunities for Innovations},'' \emph{IEEE
  Communications Magazine}, vol.~53, no.~2, pp. 90--97, 2015.

\bibitem{manzalini2013Clouds}
A.~Manzalini, R.~Minerva, F.~Callegati, W.~Cerroni, and A.~Campi, ``{Clouds of
  Virtual Machines in Edge Networks},'' \emph{IEEE Communications Magazine},
  vol.~51, no.~7, pp. 63--70, 2013.

\bibitem{bonomi2014Fog}
F.~Bonomi, R.~Milito, P.~Natarajan, and J.~Zhu, ``{Fog Computing: A Platform
  for Internet of Things and Analytics},'' in \emph{Big data and internet of
  things: A roadmap for smart environments}, 2014, pp. 169--186.

\bibitem{moreno2017cross}
R.~Moreno-Vozmediano, R.~S. Montero, E.~Huedo, and I.~M. Llorente,
  ``{Cross-site Virtual Network in Cloud and Fog Computing},'' \emph{IEEE Cloud
  Computing}, vol.~4, no.~2, pp. 46--53, 2017.

\bibitem{moreno2015beacon}
R.~Moreno-Vozmediano, E.~Huedo, I.~M. Llorente, R.~S. Montero, P.~Massonet,
  M.~Villari, G.~Merlino, A.~Celesti, A.~Levin, L.~Schour \emph{et~al.},
  ``{BEACON: A Cloud Network Federation Framework},'' in \emph{European
  Conference on Service-Oriented and Cloud Computing}, 2015, pp. 325--337.

\bibitem{yang2016seamless}
B.~Yang, W.~K. Chai, G.~Pavlou, and K.~V. Katsaros, ``{Seamless Support of Low
  Latency Mobile Applications with NFV-Enabled Mobile Edge-Cloud},'' in
  \emph{Cloud Networking (Cloudnet), 2016 5th IEEE International Conference
  on}.\hskip 1em plus 0.5em minus 0.4em\relax IEEE, 2016, pp. 136--141.

\bibitem{yang2018cost}
B.~Yang, W.~K. Chai, Z.~Xu, K.~V. Katsaros, and G.~Pavlou, ``{Cost-Efficient
  NFV-Enabled Mobile Edge-Cloud for Low Latency Mobile Applications},''
  \emph{IEEE Transactions on Network and Service Management}, 2018.

\bibitem{vinel2017emerging}
A.~Vinel, J.~Breu, T.~H. Luan, and H.~Hu, ``{Emerging Technology for 5G-enabled
  Vehicular Networks},'' \emph{IEEE Wireless Communications}, vol.~24, no.~6,
  pp. 12--12, 2017.

\bibitem{truong2015software}
N.~B. Truong, G.~M. Lee, and Y.~Ghamri-Doudane, ``Software defined
  networking-based vehicular adhoc network with fog computing,'' in
  \emph{Integrated Network Management (IM), 2015 IFIP/IEEE International
  Symposium on}, 2015, pp. 1202--1207.

\bibitem{huang2017exploring}
X.~Huang, R.~Yu, J.~Kang, Y.~He, and Y.~Zhang, ``{Exploring Mobile Edge
  Computing for 5G-enabled Software Defined Vehicular Networks},'' \emph{IEEE
  Wireless Communications}, vol.~24, no.~6, pp. 55--63, 2017.

\bibitem{ivanov2016gravity}
S.~Ivanov, S.~Balasubramaniam, D.~Botvich, and O.~B. Akan, ``{Gravity Gradient
  Routing for Information Delivery in Fog Wireless Sensor Networks},'' \emph{Ad
  Hoc Networks}, vol.~46, pp. 61--74, 2016.

\bibitem{hakiri2017managing}
A.~Hakiri, B.~Sellami, P.~Patil, P.~Berthou, and A.~Gokhale, ``{Managing
  Wireless Fog Networks using Software-Defined Networking},'' in \emph{IEEE/ACS
  14th International Conference on Computer Systems and Applications}, 2017.

\bibitem{karagiannis2015survey}
V.~Karagiannis, P.~Chatzimisios, F.~Vazquez-Gallego, and J.~Alonso-Zarate, ``{A
  Survey on Application Layer Protocols for the Internet of Things},''
  \emph{Transaction on IoT and Cloud Computing}, vol.~3, no.~1, pp. 11--17,
  2015.

\bibitem{tato2017designing}
G.~Tato, M.~Bertier, and C.~Tedeschi, ``{Designing Overlay Networks for
  Decentralized Clouds},'' in \emph{International Workshop on the Future of
  Cloud Computing and Cloud Services}, 2017.

\bibitem{aditya2017frugal}
S.~Aditya and R.~J. Figueiredo, ``{Frugal: Building Degree-Constrained Overlay
  Topology from Social Graphs},'' in \emph{IEEE 1st International Conference on
  Fog and Edge Computing}.\hskip 1em plus 0.5em minus 0.4em\relax IEEE, 2017,
  pp. 11--20.

\bibitem{chandra2013decentralized}
A.~Chandra, J.~Weissman, and B.~Heintz, ``{Decentralized Edge Clouds},''
  \emph{IEEE Internet Computing}, vol.~17, no.~5, pp. 70--73, 2013.

\bibitem{ryden2014nebula}
M.~Ryden, K.~Oh, A.~Chandra, and J.~Weissman, ``{Nebula: Distributed Edge Cloud
  for Data Intensive Computing},'' in \emph{IEEE International Conference on
  Cloud Engineering}, 2014, pp. 57--66.

\bibitem{hong2013mobile}
K.~Hong, D.~Lillethun, U.~Ramachandran, B.~Ottenw{\"a}lder, and B.~Koldehofe,
  ``{Mobile Fog: A Programming Model for Large-scale Applications on the
  Internet of Things},'' in \emph{Proceedings of the second ACM SIGCOMM
  workshop on Mobile cloud computing}, 2013, pp. 15--20.

\bibitem{tang2015hierarchical}
B.~Tang, Z.~Chen, G.~Hefferman, T.~Wei, H.~He, and Q.~Yang, ``{A Hierarchical
  Distributed Fog Computing Architecture for Big Data Analysis in Smart
  cities},'' in \emph{Proceedings of the ASE BigData \& Social
  Informatics}.\hskip 1em plus 0.5em minus 0.4em\relax ACM, 2015, p.~28.

\bibitem{nastic2017serverless}
S.~Nastic, T.~Rausch, O.~Scekic, S.~Dustdar, M.~Gusev, B.~Koteska, M.~Kostoska,
  B.~Jakimovski, S.~Ristov, and R.~Prodan, ``{A Serverless Real-Time Data
  Analytics Platform for Edge Computing},'' \emph{IEEE Internet Computing},
  vol.~21, no.~4, pp. 64--71, 2017.

\bibitem{cuervo2010maui}
E.~Cuervo, A.~Balasubramanian, D.-k. Cho, A.~Wolman, S.~Saroiu, R.~Chandra, and
  P.~Bahl, ``{MAUI: Making Smartphones Last Longer with Code Offload},'' in
  \emph{Proceedings of the 8th international conference on Mobile systems,
  applications, and services}.\hskip 1em plus 0.5em minus 0.4em\relax ACM,
  2010, pp. 49--62.

\bibitem{kosta2012thinkair}
S.~Kosta, A.~Aucinas, P.~Hui, R.~Mortier, and X.~Zhang, ``{Thinkair: Dynamic
  Resource Allocation and Parallel Execution in the Cloud for Mobile Code
  Offloading},'' in \emph{Infocom, 2012 Proceedings IEEE}, 2012, pp. 945--953.

\bibitem{jang2016soul}
M.~Jang, H.~Lee, K.~Schwan, and K.~Bhardwaj, ``{SOUL: An Edge-Cloud System for
  Mobile Applications in a Sensor-rich World},'' in \emph{IEEE/ACM Symposium on
  Edge Computing}, 2016, pp. 155--167.

\bibitem{silva2017using}
P.~M.~P. Silva, J.~Rodrigues, J.~Silva, R.~Martins, L.~Lopes, and F.~Silva,
  ``{Using Edge-Clouds to Reduce Load on Traditional Wifi Infrastructures and
  Improve Quality of Experience},'' in \emph{1st International Conference on
  Fog and Edge Computing}.\hskip 1em plus 0.5em minus 0.4em\relax IEEE, 2017,
  pp. 61--67.

\bibitem{bellavista2018human}
P.~Bellavista, S.~Chessa, L.~Foschini, L.~Gioia, and M.~Girolami,
  ``{Human-Enabled Edge Computing: Exploiting the Crowd as a Dynamic Extension
  of Mobile Edge Computing},'' \emph{IEEE Communications Magazine}, vol.~56,
  no.~1, pp. 145--155, 2018.

\bibitem{ganti2011mobile}
R.~K. Ganti, F.~Ye, and H.~Lei, ``{Mobile Crowdsensing: Current State and
  Future Challenges},'' \emph{IEEE Communications Magazine}, vol.~49, no.~11,
  2011.

\bibitem{selimi2015Cloud}
M.~Selimi, A.~M. Khan, E.~Dimogerontakis, F.~Freitag, and R.~P. Centelles,
  ``Cloud services in the guifi. net community network,'' \emph{Computer
  Networks}, vol.~93, pp. 373--388, 2015.

\bibitem{khan2017Edge}
A.~M. Khan and F.~Freitag, ``{On Edge Cloud Service Provision with Distributed
  Home Servers},'' in \emph{Cloud Computing Technology and Science (CloudCom),
  2017 IEEE International Conference on}.\hskip 1em plus 0.5em minus
  0.4em\relax IEEE, 2017, pp. 223--226.

\bibitem{sefraoui2012openstack}
O.~Sefraoui, M.~Aissaoui, and M.~Eleuldj, ``{OpenStack: Toward an Open-source
  Solution for Cloud Computing},'' \emph{International Journal of Computer
  Applications}, vol.~55, no.~3, 2012.

\bibitem{santoro2017Foggy}
D.~Santoro, D.~Zozin, D.~Pizzolli, F.~De~Pellegrini, and S.~Cretti, ``{Foggy: A
  Platform for Workload Orchestration in a Fog Computing Environment},'' in
  \emph{IEEE International Conference on Cloud Computing Technology and
  Science}, 2017, pp. 231--234.

\bibitem{vogler2015leonore}
M.~V{\"o}gler, J.~Schleicher, C.~Inzinger, S.~Nastic, S.~Sehic, and S.~Dustdar,
  ``{LEONORE--Large-Scale Provisioning of Resource-constrained IoT
  Deployments},'' in \emph{IEEE Symposium on Service-Oriented System
  Engineering}, 2015, pp. 78--87.

\bibitem{nastic2014provisioning}
S.~Nastic, S.~Sehic, D.-H. Le, H.-L. Truong, and S.~Dustdar, ``{Provisioning
  Software-Defined IoT Cloud Systems},'' in \emph{International Conference on
  Future Internet of Things and Cloud}, 2014, pp. 288--295.

\bibitem{nastic2016middleware}
S.~Nastic, H.-L. Truong, and S.~Dustdar, ``{A Middleware Infrastructure for
  Utility-based Provisioning of IoT Cloud Systems},'' in \emph{IEEE/ACM
  Symposium on Edge Computing}, 2016, pp. 28--40.

\bibitem{disc-3}
E.~Saurez, K.~Hong, D.~Lillethun, U.~Ramachandran, and B.~Ottenw\"{a}lder,
  ``{Incremental Deployment and Migration of Geo-distributed Situation
  Awareness Applications in the Fog},'' in \emph{Proceedings of the 10th ACM
  International Conference on Distributed and Event-based Systems}, 2016, pp.
  258--269.

\bibitem{disc-2}
B.~Varghese, N.~Wang, J.~Li, and D.~Nikolopoulos, ``{Edge-as-a-Service: Towards
  Distributed Cloud Architectures},'' in \emph{International Conference on
  Parallel Computing}, ser. Advances in Parallel Computing.\hskip 1em plus
  0.5em minus 0.4em\relax IOS Press, 2017, pp. 784--793.

\bibitem{disc-4}
R.~Kolcun, D.~Boyle, and J.~A. McCann, ``{Optimal processing node discovery
  algorithm for distributed computing in IoT},'' in \emph{5th International
  Conference on the Internet of Things}, 2015, pp. 72--79.

\bibitem{bench-1}
B.~Varghese, O.~Akgun, I.~Miguel, L.~Thai, and A.~Barker, ``{Cloud Benchmarking
  for Performance},'' in \emph{6th IEEE International Conference on Cloud
  Computing Technology and Science}, 2014, pp. 535--540.

\bibitem{bench-2}
J.~J. Dongarra, P.~Luszczek, and A.~Petitet, ``The linpack benchmark: past,
  present and future,'' \emph{Concurrency and Computation: practice and
  experience}, vol.~15, no.~9, pp. 803--820, 2003.

\bibitem{bench-3}
D.~H. Bailey, E.~Barszcz, J.~T. Barton, D.~S. Browning, R.~L. Carter, L.~Dagum,
  R.~A. Fatoohi, P.~O. Frederickson, T.~A. Lasinski, R.~S. Schreiber, H.~D.
  Simon, V.~Venkatakrishnan, and S.~K. Weeratunga, ``{The NAS Parallel
  Benchmarks - Summary and Preliminary Results},'' in \emph{Proceedings of the
  1991 ACM/IEEE Conference on Supercomputing}, 1991, pp. 158--165.

\bibitem{bench-4}
M.~Ferdman, A.~Adileh, O.~Kocberber, S.~Volos, M.~Alisafaee, D.~Jevdjic,
  C.~Kaynak, A.~D. Popescu, A.~Ailamaki, and B.~Falsafi, ``{Clearing the
  Clouds: A Study of Emerging Scale-out Workloads on Modern Hardware},''
  \emph{Proceedings of the Seventeenth International Conference on
  Architectural Support for Programming Languages and Operating Systems}, 2012.

\bibitem{bench-5}
T.~Palit, Y.~Shen, and M.~Ferdman, ``{Demystifying Cloud Benchmarking},'' in
  \emph{IEEE International Symposium on Performance Analysis of Systems and
  Software}, 2016, pp. 122--132.

\bibitem{bench-6}
B.~Varghese, O.~Akgun, I.~Miguel, L.~Thai, and A.~Barker, ``{Cloud Benchmarking
  For Maximising Performance of Scientific Applications},'' \emph{IEEE
  Transactions on Cloud Computing}, 2017.

\bibitem{bench-7}
B.~Varghese, L.~T. Subba, L.~Thai, and A.~Barker, ``{DocLite: A Docker-Based
  Lightweight Cloud Benchmarking Tool},'' in \emph{16th IEEE/ACM International
  Symposium on Cluster, Cloud and Grid Computing}, 2016, pp. 213--222.

\bibitem{bench-8}
Z.~Kozhirbayev and R.~O. Sinnott, ``{A Performance Comparison of
  Container-based Technologies for the Cloud},'' \emph{Future Generation
  Computer Systems}, vol.~68, pp. 175 -- 182, 2017.

\bibitem{bench-9}
R.~Morabito, ``{Virtualization on Internet of Things Edge Devices With
  Container Technologies: A Performance Evaluation},'' \emph{IEEE Access},
  vol.~5, pp. 8835--8850, 2017.

\bibitem{bench-10}
R.-A. Cherrueau, D.~Pertin, A.~Simonet, A.~Lebre, and M.~Simonin, ``{Toward a
  Holistic Framework for Conducting Scientific Evaluations of OpenStack},'' in
  \emph{Proceedings of the 17th IEEE/ACM International Symposium on Cluster,
  Cloud and Grid Computing}, 2017, pp. 544--548.

\bibitem{bench-11}
S.~Sridhar and M.~E. Tolentino, ``{Evaluating Voice Interaction Pipelines at
  the Edge},'' in \emph{IEEE International Conference on Edge Computing}, 2017,
  pp. 248--251.

\bibitem{bench-12}
A.~Krylovskiy, ``{Internet of Things Gateways Meet Linux Containers:
  Performance Evaluation and Discussion},'' in \emph{IEEE 2nd World Forum on
  Internet of Things (WF-IoT)}, 2015, pp. 222--227.

\bibitem{bench-13}
B.~Confais, A.~Lebre, and B.~Parrein, ``{Performance Analysis of Object Store
  Systems in a Fog/Edge Computing Infrastructures},'' in \emph{IEEE
  International Conference on Cloud Computing Technology and Science}, 2016,
  pp. 294--301.

\bibitem{bench-14}
M.~Ficco, C.~Esposito, Y.~Xiang, and F.~Palmieri, ``{Pseudo-Dynamic Testing of
  Realistic Edge-Fog Cloud Ecosystems},'' \emph{IEEE Communications Magazine},
  vol.~55, no.~11, pp. 98--104, 2017.

\bibitem{he2016novel}
X.~He, Z.~Ren, C.~Shi, and J.~Fang, ``{A Novel Load Balancing Strategy of
  Software-defined Cloud/Fog Networking in the Internet of Vehicles},''
  \emph{China Communications}, vol.~13, no.~2, pp. 140--149, 2016.

\bibitem{parsopoulos2002particle}
K.~E. Parsopoulos, M.~N. Vrahatis \emph{et~al.}, ``{Particle Swarm Optimization
  Method for Constrained Optimization Problems},'' \emph{Intelligent
  Technologies--Theory and Application: New Trends in Intelligent
  Technologies}, vol.~76, no.~1, pp. 214--220, 2002.

\bibitem{beraldi2017cooperative}
R.~Beraldi, A.~Mtibaa, and H.~Alnuweiri, ``{Cooperative Load Balancing Scheme
  for Edge Computing Resources},'' in \emph{2nd International Conference on Fog
  and Mobile Edge Computing}.\hskip 1em plus 0.5em minus 0.4em\relax IEEE,
  2017, pp. 94--100.

\bibitem{ningning2016Fog}
S.~Ningning, G.~Chao, A.~Xingshuo, and Z.~Qiang, ``{Fog Computing Dynamic Load
  Balancing Mechanism Based on Graph Repartitioning},'' \emph{China
  Communications}, vol.~13, no.~3, pp. 156--164, 2016.

\bibitem{puthal2018secure}
D.~Puthal, M.~S. Obaidat, P.~Nanda, M.~Prasad, S.~P. Mohanty, and A.~Y. Zomaya,
  ``{Secure and Sustainable Load Balancing of Edge Data Centers in Fog
  Computing},'' \emph{IEEE Communications Magazine}, vol.~56, no.~5, pp.
  60--65, 2018.

\bibitem{dastjerdi2016Fog}
A.~V. Dastjerdi, H.~Gupta, R.~N. Calheiros, S.~K. Ghosh, and R.~Buyya, ``Fog
  computing: Principles, architectures, and applications,'' in \emph{Internet
  of Things}.\hskip 1em plus 0.5em minus 0.4em\relax Elsevier, 2016, pp.
  61--75.

\bibitem{wang2017dynamic}
S.~Wang, R.~Urgaonkar, T.~He, K.~Chan, M.~Zafer, and K.~K. Leung, ``Dynamic
  service placement for mobile micro-clouds with predicted future costs,''
  \emph{IEEE Transactions on Parallel and Distributed Systems}, vol.~28, no.~4,
  pp. 1002--1016, 2017.

\bibitem{taneja2017resource}
M.~Taneja and A.~Davy, ``Resource aware placement of iot application modules in
  fog-cloud computing paradigm,'' in \emph{Integrated Network and Service
  Management (IM), 2017 IFIP/IEEE Symposium on}.\hskip 1em plus 0.5em minus
  0.4em\relax IEEE, 2017, pp. 1222--1228.

\bibitem{gupta2017iFogsim}
H.~Gupta, A.~Vahid~Dastjerdi, S.~K. Ghosh, and R.~Buyya, ``ifogsim: A toolkit
  for modeling and simulation of resource management techniques in the internet
  of things, edge and fog computing environments,'' \emph{Software: Practice
  and Experience}, vol.~47, no.~9, pp. 1275--1296, 2017.

\bibitem{skarlat2017towards}
O.~Skarlat, M.~Nardelli, S.~Schulte, and S.~Dustdar, ``Towards qos-aware fog
  service placement,'' in \emph{Fog and Edge Computing (ICFEC), 2017 IEEE 1st
  International Conference on}.\hskip 1em plus 0.5em minus 0.4em\relax IEEE,
  2017, pp. 89--96.

\end{thebibliography}

\end{document}